%% file: fibonacci.tex
\documentclass[runningheads]{llncs}
\usepackage[english]{babel}
\usepackage[T1]{fontenc}
\usepackage[utf8]{inputenc}
\usepackage{cancel}
\usepackage[obeyspaces]{url}
\newcommand{\comm}[1]{}

\usepackage{mathrsfs, frcursive, microtype, mathtools, amsmath,
  amssymb, mathtools, tikz, tikz-cd, subcaption, framed, float,
  tensor, bm, dsfont, floatpag, mathabx, microtype, multirow}

\usepackage[normalem]{ulem}

\usepackage[
  bookmarks=false, 
  colorlinks,
  citecolor=black,
  linkcolor=black,
  urlcolor=black,
]{hyperref}
\usepackage{cleveref}
\usetikzlibrary{shapes.geometric, backgrounds, calc,
  arrows, automata, fit, positioning, patterns,
  decorations.pathreplacing, tikzmark, decorations.markings}

\definecolor{MyRose}{HTML}{DC476D}
\definecolor{MyGreen}{HTML}{7BA35F}
\definecolor{sk}{HTML}{BD0D02}
\definecolor{forestgreen}{rgb}{0.13, 0.55, 0.13}

\newcommand{\bb}[1]{\bm{\mathrm{#1}}}

\newcommand\oeis[1]{\href{https://oeis.org/#1}{#1}}

\newcommand\reverse[1]{\protect\overleftarrow{#1}}

\DeclareMathOperator{\first}{first}
\DeclareMathOperator{\last}{last}

\def\ONE{\mathds{1}}
\def\S{\mathcal{S}}

\newcommand{\Bw}{\mathcal{B}}

\newcommand{\W}{\mathcal{W}}

\author{Jean-Luc Baril \and
Sergey Kirgizov \and
Vincent Vajnovszki
}

\authorrunning{J.-L. Baril, S. Kirgizov and V. Vajnovszki}

\institute{
  LIB, Université de Bourgogne Franche-Comté \\
  B.P. 47 870, 21078 Dijon Cedex France \\
\email{\{barjl,sergey.kirgizov,vvajnov\}@u-bourgogne.fr}}

\title{Gray codes for Fibonacci $q$-decreasing words}
\begin{document}
\newtheorem{conj}{Conjecture}
\newtheorem{NOTE}{Note}
\maketitle

\begin{abstract}

A length $n$ binary word is $q$-decreasing, $q\geq 1$, if every of its
maximal factors of the form $0^a1^b$ satisfies $a=0$ or $q\cdot a >
b$.  We show constructively that these words are in bijection with
binary words having no occurrences of $1^{q+1}$, and thus they are
enumerated by the $(q+1)$-generalized Fibonacci numbers.  We give some
enumerative results and reveal similarities between $q$-decreasing
words and binary words having no occurrences of $1^{q+1}$ in terms of
frequency of $1$ bit.  In the second part of our paper, we provide an
efficient exhaustive generating algorithm for $q$-decreasing words in
the lexicographic order, for any $q\geq 1$, show the existence of
3-Gray codes and explain how a generating algorithm for these Gray
codes can be obtained.  Moreover, we give the construction of a more
restrictive 1-Gray code for $1$-decreasing words, which in particular
settles a conjecture stated recently in the context of interconnection
networks by Eğecioğlu and Iršič.
\end{abstract}

\section{Introduction and preliminaries}
\label{int}

The Fibonacci sequence origins have been traced back to the works of
ancient Indian mathematician \={A}c\={a}rya Pi\.{n}gala dealing with
rhythmic structure patterns in Sanskrit
poetry~\cite[p. 50]{singh,knuth43}. Over time, the study of words and
patterns became more abstract and systematic (see for instance
Lothaire's books~\cite{l1,l2,l3} and~\cite{Berstel}).  A considerable
amount of questions concerning efficient enumeration and generation of
words respecting certain properties (including pattern avoidance) were
mathematically formulated and answered only relatively recently, the
works closely related to the present study
include~\cite{barilucas,bernini,ege1,ege2,gray,squire,vaj-fib,lin}.

In this paper we introduce $q$-decreasing words, a novel class of
run-restricted binary words enumerated by the $(q+1)$-generalized
Fibonacci numbers, $q\geq 1$.  For $q=1$ the subclass of such words
that start with $0$ was recently considered in the context of induced
subgraphs of hypercubes~\cite{ege1,ege2}.  The
  $q$-decreasing words could be interesting objects to study in other
  domains, for example in stringology, the study of string
  algorithms~\cite{stringology}. In Section~\ref{sbij} we present a
bijection between this novel class of words and Fibonacci words,
i.e. binary words avoiding consecutive~$1$s.  Section~\ref{genfun} is
devoted to the presentation of several generating functions and
enumeration results.  Finally, in Section~\ref{grayc0de}, we show the
existence of a 3-Gray code for any $q\geq 1$, give an efficient
exhaustive generating algorithms and a much more involved construction
for a 1-Gray code in the special case $q=1$. In particular, the latter
Gray code gives a Hamiltonian path in Fibonacci-run graphs whose
existence is conjectured in~\cite{ege1}.

\bigskip

The following set of notations is adopted. Let $\Bw$ denote the set of
all finite-length binary words, i.e. strings over the alphabet
$\{0,1\}$, and $\Bw_n$, $n \ge 0$, be the set of all binary words of
length $n$.  For a given binary word $w$ we use the notation $w_i$ to
mean the letter at position $i$.

A nonempty sequence of adjacent
letters inside a word is called {\em factor}.  A factor $v$ repeated
$k$ times is denoted by $v^k$, for instance $(00)^21^2 = 000011$.  For
a given length $n$, the notation $v^*$ is used to repeat factor $v$ as
many times as possible, until the length $n$ is reached, possibly
trimming extra symbols at the end; and the length $n$ will be understood
from the context.  For example, if a word $v$ of length $n=7$ is equal
to $(001)^*$ it means $v = 0010010$.

The set of all $n$-length binary words containing no occurrences of
factor $v$ is denoted by $\Bw_n(v)$.
Let $\Bw(v) = \bigcup_{n = 0}^{\infty}\Bw_n(v)$.
The concatenation of two words
$w$ and $v$ is denoted by $w \cdot v$ or simply by $w v$.  If $v$ is a
binary word and $\mathcal{W}$ is a set of binary words, let
$\mathcal{W} \cdot v = \{ w \cdot v \mid w \in \mathcal{W} \}$, and $v
\cdot \mathcal{W}$ is defined similarly.  Whenever $\mathcal{A}$ and
$\mathcal{C}$ are two subsets of $\Bw$, we define $\mathcal{A} \cdot
\mathcal{C} = \{ a \cdot c \mid a \in \mathcal{A}, c \in \mathcal{C}\}$.

\medskip

Following~\cite{miles} the $n$th $k$-generalized Fibonacci number is defined as

\begin{equation}
f_{n,k} = \begin{cases}
0 & \text{if } 0 \leq n \leq k - 2,\\
1 & \text{if }  n = k - 1,\\
\sum_{i=1}^k f_{n-i,k} & \text{otherwise}.
\end{cases}
\label{fib_seq}
\end{equation}

  As noted in~\cite{vaj-fib}, the generating function $g_k(x) = \sum_{n=0}^\infty
  f_{n,k} x^n$
  for
  $k$-generalized Fibonacci numbers is
  \begin{equation}
    g_k(x) = \frac{x^{k-1}}{1-x-x^2-\cdots-x^k} = \frac{x^{k-1} - x^k}{1-2x+x^{k+1}}.
    \label{fib_classical_gf}
  \end{equation}
Related constructions also appear in~\cite[p. 42]{book} and
in~\cite[p. 309]{Jorg_ardnt}.

\noindent{\bf Classical fact.} 
{\em The number of words in $\Bw_n(1^k)$ equals $f_{n+k,k}$ for $k\ge 2$, moreover}
\begin{equation} 
\Bw_n \big(1^k\big) =
\begin{cases}
    \Bw_n & \text{if } n<k,\\
    \bigcup_{i=0}^{k-1}\, 1^i0 \cdot \Bw_{n-i-1} \big( 1^k \big) & \text{otherwise.}
\end{cases}
\label{decoposition}
\end{equation}

The classical fact comes, for instance, from~\cite[p. 286]{knuth3}.
The binary words avoiding two consecutive 1s are counted by Fibonacci
numbers, words without factor 111 are counted by Tribonacci numbers,
etc.  We call such words {\em (generalized) Fibonacci words}.  The
On-line Encyclopedia of Integer Sequences founded by
N.J.A. Sloane~\cite{sloane} contains several corresponding sequences
(see for example \oeis{A000045} and \oeis{A000073}, after taking a
binary complement). Gray codes for Fibonacci words are discussed  in~\cite{vaj-fib}, and more generally, Gray codes for words avoiding a given factor in~\cite{BBPSV,squire}.

\begin{lemma}
  For $q \ge 1$, the bivariate generating function $B_q(x,y)=
  \sum_{n,k\ge 0} b_{n,k} x^ny^k$ where the coefficient $b_{n,k}$ of
  $x^ny^k$ is the number of Fibonacci words of length $n$ having $k$
  $1$s in $\Bw_n(1^{q+1})$ is
  $$B_q(x,y)=\frac{y\left(1-(xy)^{q+1}\right)}{y-xy^2-xy+(xy)^{q+2}}.$$
  \label{classical_lemma}
  \end{lemma}

\proof
Alternative to relation (\ref{decoposition}), the set $\Bw(1^{q+1})$ of (any
length) binary words avoiding $1^{q+1}$ can be defined recursively as
$$\Bw(1^{q+1})=\ONE_q\cup \bigcup_{i=0}^q 1^i0\cdot \Bw(1^{q+1})$$
where $\ONE_q = \bigcup_{i=0}^q \{1^i\}$. It follows that the
bivariate generating function $B_q(x,y)$ satisfies the functional equation
$$B_q(x,y)=\sum_{i=0}^qx^iy^i+ B_q(x,y)\sum_{i=0}^{q}x^{i+1}y^{i},$$
and we have
$B_q(x,y)=\frac{y\left(1-(xy)^{q+1}\right)}{y-xy^2-xy+(xy)^{q+2}}$.
\qed
\endproof

It is not surprising that $B_q(x,1)=\frac{g_{q+1}(x)-x^q}{x^{q+1}}$, see relation \eqref{fib_classical_gf}.
Indeed, both sides of this equality are generating functions for the sequence $(f_{n+q+1,q+1})_{n\geq
 0}$ counting words in $\Bw(1^{q+1})$, which is the  $(q+1)$th left shift of the sequence $(f_{n,q+1})_{n\geq
0}$, see relation \eqref{fib_seq} and the classical fact following it.


\medskip

The {\em Hamming distance} between two binary words of the same length
equals the number of positions at which they differ.  A $k$-Gray code
for a set $\mathcal{A} \subset \Bw_n$ is an ordered list $\bb{A}$ for
$\mathcal{A}$, such that the Hamming distance between any two
consecutive words in $\bb{A}$ is at most $k$, and we say that words in
$\bb{A}$ are listed in {\em Gray code order}.  Frank Gray's
patent~\cite{gray} discusses an early example and application of such
a code for the set of $n$-length binary words. The concatenation of
two ordered lists $\bb{L}_1$ and $\bb{L}_2$ is denoted by $\bb{L}_1
\circ \bb{L}_2$, and $\reverse{\bb{L}}$ designates the reverse of the
list $\bb{L}$.  If $\bb{L}$ is a list of words, then
$\langle\bb{L}\rangle^i = \bb{L}$ whenever $i$ is even,
  and $\langle\bb{L}\rangle^i = \reverse{\bb{L}}$ otherwise.
First and last elements of $\bb{L}$ are denoted respectively by
$\first(\bb{L})$ and $\last(\bb{L})$. Also, we denote by
$\widehat{\bb{L}}$ the list obtained from $\bb{L}$ by deleting its
first element. 

  For example, a list containing elements 00, 01 and 11
  will be noted as $\bb{L} = 00 \circ 01 \circ 11$, and
  $\widehat{\bb{L}} = 00 \circ 01$.

\comm{
\begin{definition}
 A word whose run-length representation 
 $0^{r_1}1^{r_2}0^{r_3}\ldots
 0^{r_{m-1}}1^{r_m}$ where $m \ge 2, r_1 \ge 0, r_m \ge 0$
 and $r_i \ge
 1$ for any $i \notin \{1,m\}$ respects the following 
every odd $i$, will be called {\em $q$-decreasing}:
$$r_i > 0 \Rightarrow r_i > r_{i+1}/q.$$
\end{definition}
}
\begin{definition}
A binary word is called {\em $q$-decreasing}, for
$ q \geq 1$,
if any of its length maximal factors of the 
form $0^a1^b$, $a>0$, satisfies $q\cdot a > b$.
\end{definition}

The set of $q$-decreasing words of length $n$ is denoted by $\W^q_n$.
For example we have 
$\W^1_4 = \{ 0000, 0001, 0010, 1000, 1001, 1100, 1110, 1111 \}$. See also Table \ref{example_Gray} for 
the sets $\W^2_4$ and $\W^1_6$.
Let $\W^q = \bigcup_{n = 0}^{\infty}\W^q_n$.

\section{Bijection with classical Fibonacci words}
\label{sbij}

In this section we prove that $q$-decreasing words, $q\geq 1$, are
enumerated by $(q+1)$-generalized Fibonacci numbers defined in 
relation (\ref{fib_seq}). We start with a definition and several
propositions.

\begin{definition}
  For any $q \ge 1$, we define the map $\psi^q$ from $\Bw_n$ to
  $\Bw_{n+q+1}$ as

  \normalfont
  $$
    \psi^q (w) = \begin{cases}
      v001^{k+q} & \text{ if } w = v 0 1^k, v \in \Bw, k \ge 0,\\
      1^{n+q+1}  & \text{ otherwise. }
    \end{cases}
  $$ 
\end{definition}

Less formally, $\psi^q$ inserts a factor $01^q$ immediately after the
last occurrence of $0$, and it adds the suffix $1^{q+1}$ to the word
containing no $0$. For example $\psi^1(0) = 0 {\bf 01}$, $\psi^1(00011) =
000 {\bf 01} 11$, $\psi^2(0011101) = 001110 {\bf 011} 1$ and $\psi^5(1) =
1{\bf 111111}$. The value of $q$ will be clear from the context, so by
slight abuse of notation $\psi^q$ will be denoted $\psi$ throughout
the paper.

\begin{proposition}
 For $n\ge 0, q\ge 1$, $\psi$ is an injective map from from $\Bw_n$
  to $\Bw_{n+q+1}$.\label{psi1}
\end{proposition}
\proof For two $n$-length words $w \ne w'$ we show that $\psi(w) \ne \psi(w')$.
  It is clear that if one of the given words contains no $0$
  the injectivity holds.  Otherwise we have two cases. If $w = v 01^k$
  and $w' = v' 01^k$ then we have necessarily $v \ne v'$ and $v001^{k+q} \ne
  v'001^{k+q}$, so the images are different.  If $w = v 01^k$ and $w'
  = v' 01^\ell$ with $ k \neq \ell$, then $ v001^{k+q} \ne
  v'001^{\ell+q}$ and again $\psi(w) \ne \psi(w')$.
  \qed
\endproof

In the following, we will use the restriction of $\psi$ to the set of
$q$-decreasing words, namely $\psi : \W^q_n \to \W^q_{n+q+1}$. It is
possible due to Proposition~\ref{psi-rest} below.

\begin{proposition} 
For $n \ge 0, q \ge 1$, $\psi(\W^q_n)$ consists of all $q$-decreasing words of length $n+q+1$ ending with at least $q$ ones.
\label{psi-rest}
\end{proposition}
\proof
If $w=1^n$, then $\psi(w)=1^{n+q+1}$.
Otherwise, we write $w= v 0^a1^b$ where $a > b/q\ge 0$ and the word $v$ is either empty or ends with 1. 
So $\psi (v0^a 1^b) = v 0^{a+1}1^{q+b}$.  As we have $1 + a > 1 + b/q = (q+b)/q$, $\psi(w)$ is a $q$-decreasing word ending with at least $q$ $1$s.  
Similarly, any $(n+q+1)$-length $q$-decreasing word ending with at least $q$ $1$s
can be obtained from a (unique) word in $\W^q_n$ by~$\psi$.
\qed
\endproof

Now, we present a one-to-one correspondence between Fibonacci and
$q$-decreasing words. Recall that, for $q\geq 1$, the set
$\Bw_n(1^{q+1})$ of $(q+1)$-generalized Fibonacci words is the set of
binary words of length $n$ with no $1^{q+1}$ factors, see relation (\ref{decoposition}) for the recursive definition of these words
according to their length.


\begin{definition}
 We define the map $\phi
 :\Bw_n(1^{q+1})\to \Bw_n$ as \normalfont \def\mmmaps{\xmapsto{\quad \phi
     \quad}}
\begin{equation}
  \phi(w) =
  \begin{cases}
      1^k & \text{ if } w= 1^k \text{ and } k \in [0, q], \\
      \psi \big( \phi (v)  \big) & \text{ if } w=1^q0 v,  \\
      \phi(v) 01^{k} & \text{ if } w = 1^k0  v  \text{ and }  k \in [0, q-1].
  \end{cases}
\label{def-phi}
\end{equation}

  \label{bij-def}
\end{definition}
\noindent
See Table \ref{example_Gray}(a) for the images of the words in
$\Bw_4(111)$ through $\phi$.


\begin{theorem}
For $n \ge 0, q \ge 1$,  $\phi$ maps bijectively $\Bw_n(1^{q+1})$
into $\W_n^q$.
%
\label{bij}
\end{theorem}

\proof We proceed by induction on $n$.  
%
%
The classical decomposition in
relation (\ref{decoposition}) gives rise to the three cases
specified in relation (\ref{def-phi}), and we have:\\
(i) If  $n\leq q$, then the word $1^n$ is sent by $\phi$ to $1^n$; 
\\
(ii) Words of the form $1^q0v$, where $v \in
\Bw_{n-q-1}(1^{q+1})$, are sent to words ending by at least $q$ $1$s
(see Proposition ~\ref{psi-rest});\\
(iii) Words of the form $1^k0v$, where $k \in [0,
  q-1]$ and $v \in \Bw_{n-k-1}(1^{q+1})$, are sent to words ending
by at most $q-1$ $1$s.

\noindent
Using the bijectivity of $\psi$ (see
Propositions~\ref{psi1} and~\ref{psi-rest}) and the induction
hypothesis, it is routine to check that $\phi(w)\in \W_n^q$ for any $w \in \Bw_n(1^{q+1})$, and $\phi(w) \ne \phi(w')$ for any two different words $w,w'\in \Bw_n(1^{q+1})$.

Similarly, by induction on $n$, any word in $\W_n^q$ can be obtained by
$\phi$ from a word in $\Bw_n(1^{q+1})$, and the statement holds.  \qed
\endproof

It follows that  $(q+1)$-order Fibonacci words and $q$-decreasing words are
equinumerous.

\section{Some enumeration results}
\label{genfun}

Here we provide a bivariate generating function $W_q(x,y) = \sum_{n,k
  \ge 0} w_{n,k} x^ny^k$, where $w_{n,k}$ is the number of $n$-length
$q$-decreasing words having $k$ $1$s.  This bivariate generating
function is of a particular interest since it will help us (see
Corollary~\ref{parity}) to prove that $q$-decreasing words satisfy a
necessary condition for the existence of 1-Gray code, called {\em
  parity condition}. More precisely, if a set $\mathcal{A}$ of binary
words admits a 1-Gray code, and $\mathcal{A}^+$
(resp. $\mathcal{A}^-$) denotes the subset of $\mathcal{A}$ having
even (resp. odd) number of $1$s, then the {\em parity difference}
$|\mathcal{A}^+| - |\mathcal{A}^-|$ must be equal to either $0$, $1$,
or $-1$.
Indeed, the graph where the vertex set is $\mathcal{A}$ and edges connect
vertices with Hamming distance one is bipartite with partite sets $\mathcal{A}^+$ and $\mathcal{A}^-$. A Hamiltonian path in this graph (or equivalently a 1-Gray code for $\mathcal{A}$) cannot exist unless it satisfies the parity condition.
This parity condition is used for instance in~\cite{squire}
to investigate the possibility of 1-Gray code for a set of words
avoiding a given factor.

\medskip

In order to derive the expression of $W_q(x,y)$, we use the following
decomposition of the set $\W^q$:
$$
\W^q = \ONE \cup \W^q \cdot \S^q,
$$ where $\ONE = \cup_{n=0}^\infty \{1^n\}$ and $\S^q$ corresponds to
the set of all factors of the form $0^a1^b$ respecting $q$-decreasing
property (i.e. $a > b/q \ge 0$) such that none of them is a
concatenation of other factors respecting $q$-decreasing
property. More precisely, $a$ is the smallest integer strictly greater
than $b/q$, i.e. $a = \lfloor b/q \rfloor + 1$. A factor from $\S^q$
will be called {\em $q$-prime factor}, and thus $\S^q$ is the set of
such factors.  For instance: $\S^1 = \{ 0, 001, 00011, 0000111, \ldots
\}$, $\S^2 = \{ 0, 01, 0011, 00111, 0001111, 00011111, \ldots \}$.

\begin{lemma}
  For $q \ge 1$, the bivariate generating function $S_q(x,y)= \sum_{n,k\ge 0} s_{n,k}
  x^ny^k$ where the coefficient $s_{n,k}$ is the number of $q$-prime
  factors of length $n$ having exactly $k$ $1$s is:
  $$ S_q(x,y) = \frac{ x \left( 1 - (xy)^q \right)}{ (xy - 1) (x^{q+1} y^q - 1)}.$$
  \label{primes}
\end{lemma}

\proof
  Any $q$-prime factor is of the form $0^a1^b$ with $a = \lfloor b/q
  \rfloor + 1$. So, if $b = kq + r$ with $k \geq 0$ and $r \in [0,
    q-1]$, then $a + b = k(q+1) + r + 1$ and we can write:
  $$ S_q(x,y) = \sum_{k=0}^\infty \sum_{r=0}^{q-1 } x^{k(q+1) + r + 1}
  y^{kq + r}.$$ A simple calculation results in the claimed formula.
  \qed
\endproof

\begin{theorem}
  For $q \ge 1$, the bivariate generating function $W_q(x,y) =
  \sum_{n,k\ge 0} w_{n,k} x^ny^k$ where the coefficient $w_{n,k}$ is
  the number of $n$-length $q$-decreasing words containing exactly $k$
  $1$s is given by:
  $$
  W_q(x,y) = \frac{ 1 - x^{q + 1} y^q}{ 1 - x y - x + x^{q + 2} y^{q + 1}}.
  $$
  \label{biv}
\end{theorem}

\proof
Due to the decomposition $\W^q = \ONE \cup \W^q \cdot \S^q$, we
have $W_q(x,y)=\frac{1}{1-xy}\cdot\frac{1}{1-S_q(x,y)}$, and the
result holds after applying Lemma~\ref{primes}.
\qed \endproof

\begin{corollary}
For any $n \ge 0, q\geq 1$, the set $\W^q_n$ satisfies the parity condition.
  \label{parity}
\end{corollary}

\proof
The generating function $D_q(x) = \sum_{n\ge 0} d_nx^n$ where the coefficient $d_n$ is the parity difference corresponding to the set $\W^q_n$ is obtained by making the substitution $y = -1$ in $W_q(x,y)$:
  $$ D_q(x)= \frac{ (-1)^q x^{q + 1} - 1}{ (-1)^q x^{q + 2} - 1}. $$
  When $q$ is even, $ D_q(x)= \frac{ x^{q + 1} - 1}{ x^{q + 2} - 1} =
  \sum_{n=0}^\infty \big( x^{n(q+2)} - x^{n(q+2) + q + 1} \big)$,
  otherwise $ D_q(x) = \frac{ x^{q + 1} + 1}{ x^{q + 2} + 1} = \sum_{n
    = 0}^\infty (-1)^n \big( x^{n(q+2)} + x^{n(q+2) + q + 1} \big)$.
  All involved coefficients are from $\{-1, 0, 1\}$, and the parity
  condition holds.
  \qed \endproof

\medskip

The following two corollaries are obtained by respectively calculating the
expressions: $ W_q(x,1)$, $ \frac{\partial
  W_q(x,y)}{\partial y} \big\vert_{y = 1} $ and $ \frac{\partial
  W_q(xy,1/y)}{\partial y} \big\vert_{y = 1} $.

\begin{corollary}
  For $q \ge 1$, the generating function $F_q(x) = \sum_{n\ge 0} f_{n}
  x^n$ where the coefficient $f_{n}$ is the number of $n$-length
  $q$-decreasing words is given by:
  $$ F_q (x) = \frac{1 - x^{q + 1}}{1 - 2 x + x^{q + 2}}. $$
\end{corollary}
Note that, as predicted by Theorem \ref{bij}, 
$F_q(x)=B_q(x,1)$, see 
Lemma \ref{classical_lemma} and the remark following it.

The {\it popularity} of a symbol in a set of words is the overall
number of the occurrences of the symbol in the words of the set.

\begin{corollary}
  For $q \ge 1$, the generating function $P_{q,1}(x) = \sum_{n \ge 0}
  p_{n} x^n$ where the coefficient $p_{n}$ is the popularity of $1$s
  in all $n$-length $q$-decreasing words is
  $$
  P_{q,1} (x) = 
  \frac{x \left(1 - q x^{q} + q x^{q + 1} - 2 x^{q + 1} + x^{2 q +
      2} \right)}{\left(1 - 2 x + x^{q + 2} \right)^{2}}.
  $$
Similarly, the generating function for the popularity of $0$s  in all $n$-length $q$-decreasing words is
  $$ P_{q,0} (x) = \frac{x \left(1 - x^{q} \right)}{\left(1 - 2 x +
    x^{q + 2} \right)^{2}}.
$$
\label{pop10}
\end{corollary}

The
popularity of 1s in $\Bw_n(11)$ is
equal to the number 
of edges in the Fibonacci cube~\cite{hsu} of 
order $n$, see~\cite{klav} and comments to the 
sequence~\oeis{A001629} in~\cite{sloane}. The 
generating function $P_{1,0}(x)$ 
allows us to show that 
the popularity of $0$s in $\W^1_n$ is a
shift of the sequence \oeis{A006478} 
enumerating the number of edges in the {\em Fibonacci 
hypercube}~\cite{ris-cos},
i.e. in a polytope determined by the convex hull of the Fibonacci cube.

\medskip

Despite the $q$-decreasing words and Fibonacci words have quite
different definitions, they are equinumerous and share some common
features. We end this section by showing that the 1s frequency
(defined formally below) of both sets have the same limit when $n$
tends to infinity.

If $\omega_n$ (resp. $\beta_n$) is the ratio between the popularity of
$1$s and that of $0$s in the words in $\W^1_n$ (resp. in $\Bw_n(11)$),
then $\lim_{n\to \infty} \omega_n = \lim_{n\to \infty}
\beta_n$. Indeed, extracting the coefficients of $x^n$ in both
$P_{1,1}$ and of $P_{1,0}$, their ratio tends to $2-\varphi\approx
0.3819660113$ when $n$ tends to infinity, where $\varphi$ is the
golden ratio; and this is also the limit of $\beta_n$, which is
obtained by investigating the ratio of the coefficients of $x^n$ in $
\frac{\partial B_q(x,y)}{\partial y} \big\vert_{y = 1} $ and in $
\frac{\partial B_q(xy,1/y)}{\partial y} \big\vert_{y = 1} $, where
$B_q(x,y)$ is from Lemma~\ref{classical_lemma}.

The {\it 1s frequency} of a set of binary words is the ratio between
the popularity of $1$s and the overall number of bits in the words of
the set. Alternatively, it is the expected value when a bit is
randomly chosen in the words of the set.  With the notations above, we
have that the 1s frequency of $\W^1_n$ is $\frac{1}{1+\frac{1}{\omega_n}}$
and that of $\Bw_n(11)$ is $\frac{1}{1+\frac{1}{\beta_n}}$, and we have
the next result.

\begin{corollary} 
The  1s frequency of $\W^1_n$ and of $\Bw_n(11)$ both tend to $\frac{2-\varphi}{3-\varphi}$ when $n$ tends to infinity, where $\varphi$ is $\frac{1+\sqrt{5}}{2}$.
\label{limit_q=1}
\end{corollary} 

\comm{
{\color{sk}
\begin{corollary} The number of 
edges in the Fibonacci hypercube of order $n$
is asymptotically equivalent to $n \cdot v - e$, where
$v$ (resp. $e$) is the number of vertices (resp. edges) in the Fibonacci cube of order $n$.
\end{corollary}
}}

More generally, for any $q\geq 1$, the overall number of bits in both
sets $\W^q_n$ and $\Bw_n(1^{q+1})$ is $n\cdot f_{n+q+1,q+1}$, and due
to the second rule in relation (\ref{def-phi}) defining the
bijection $\phi :\Bw(1^{q+1})\to \W^q$ we have that in $\W^q_n$ there
are more $1$s than in $\Bw_n(1^{q+1})$. However, the next corollary
shows that the difference between the 1s frequency of $\W^q_n$ and
that of $\Bw_n(1^{q+1})$ tends to zero when $n$ tends to infinity.

\begin{corollary} For any $q\geq 1$, if $u_{n,q}$ (resp. $v_{n,q}$)  is the popularity of $1s$ in $\W^q_n$ (resp. in $\Bw_n(1^{q+1})$), then we have $$\lim_{n\to \infty}\frac{u_{n,q}-v_{n,q}}{n\cdot f_{n+q+1,q+1}}=0.$$
\label{limit_difference}
\end{corollary}

\proof
Since, for any $q\geq 1$, $u_{n,q}-v_{n,q}\geq 0$, it suffices to
prove that we have $u_{n,q}-v_{n,q}\leq f_{n+q+1,q+1}$.
Using Corollary~\ref{pop10}
and Lemma~\ref{classical_lemma},
  the generating function $H(x)$ where the
coefficient of $x^n$ is $ f_{n+q+1,q+1}+v_{n,q+1}-u_{n,q}$ is

\begin{eqnarray*}
H(x) & = & B_{q}(x,1)+\frac{\partial
  B_{q}(x,y)}{\partial y} \bigg\vert_{y=1}-P_{q,1}(x)\\
   & = &\frac{1-2x^{q+1}}{1-2x+x^{q+2}},
\end{eqnarray*}
which satisfies the functional equation
$H(x)=1-2x^{q+1}+2xH(x)-x^{q+2}H(x)$. By a simple observation, $H(x)$
is also the generating function with respect to the length of binary
words different from $0^{q+1}$ and $1^{q+1}$ and that do not start
with $0^{q+2}$. Then we have $u_{n,q}-v_{n,q}\leq
f_{n+q+1,q+1}$. Dividing by $nf_{n+q+1,q+1}$, and taking the limit
when $n$ tends to infinity, we obtain the expected result. \qed
\endproof

Corollary~\ref{limit_q=1} says that, for $q=1$, the 1s frequency of $\W^q_n$ and that of $\Bw_n(1^{q+1})$ have a common limit when $n$ tends to infinity. 
For $q\geq 2$, Corollary~\ref{limit_difference} does not ensure that each of the 1s frequency of $\W^q_n$ (that is $\frac{u_{n,q}}{n\cdot f_{n+q+1,q+1}}$) and that of $\Bw_n(1^{q+1})$ (that is $\frac{v_{n,q}}{n\cdot f_{n+q+1,q+1}}$) has a limit when $n$ tends to infinity. However, using asymptotic analysis (see for instance \cite{book}) it can be shown that $\frac{v_{n,q}}{n\cdot f_{n+q+1,q+1}}$ converges to a non-zero value
when $n$ tends to infinity,
and the limit can be approximated by numerical methods. From Corollary~\ref{limit_difference} it follows that so does $\frac{u_{n,q}}{n\cdot f_{n+q+1,q+1}}$, and the two limits are equal.

Since the proof of this result is beyond the scope of the present paper we state it (including the case $q=1$ in Corollary~\ref{limit_q=1}) without proof.
\begin{NOTE} 
For $q\geq 1$ the 1s frequency of $\W^q_n$ and of $\Bw_n(1^{q+1})$ have a common non-zero limit when $n$ tends to infinity.
\label{limit_q>1}
\end{NOTE}

\section{Exhaustive generation and Gray codes for $q$-decreasing words}
\label{grayc0de}
Here we show that $q$-decreasing words can be efficiently generated in
the lexicographical order and explain how the obtained generating
algorithm can be turned into a $3$-Gray code generating one. Then, we
give a more intricate construction of a $1$-Gray code for the
particular case $q=1$. As a byproduct, this construction gives a
positive answer for the existence a Hamiltonian path in Fibonacci-run
graph conjectured in~\cite{ege1}.

\subsection{$3$-Gray codes and exhaustive generation}

Algorithm in Figure~\ref{gen_alg} generates prefixes of $q$-decreasing
words in the lexicographical order, and eventually all $n$-length
$q$-decreasing words.  The size $n$, parameter $q$ and the array $w$
of length $n+1$ are global variables and the main call is {\sc
  LexFib}($1$, $n$). For convenience, $w[0]$ is initialized by $1$ and
the parameter $delta$ is the number of consecutive $1$s that can be
added to the current generated prefix without violating the $q$-decreasingness.

It can be seen that this recursive algorithm satisfies  
Baronaigien and Ruskey's {\it constant amortized time} 
({\it CAT}) {\it principle} in \cite{Bar-Rus} stated below.
As noticed in~\cite{Bar-Rus}, by considering the underlying computation tree it follows that any algorithm satisfying it is efficient, and thus so is {\sc LexFib}. We refer the reader to~\cite[Section 1.7]{Ruskey3} for more about CAT exhaustive generating algorithms.

\medskip
\noindent

{CAT principle} \cite{Bar-Rus}: A recursive generation algorithm with the following properties runs in constant amortized time. (i) Every call results in the output of at least one 
object;
(ii)~Excluding the computation done by recursive calls the amount of computation of any call is proportional to the degree of the call;
(iii)~The number of calls of degree one is $O(N)$, where $N$ is the number of generated objects.

\begin{figure}[!ht]
\begin{center}
\begin{tabular}{|c|}
\hline
\begin{minipage}[c]{.46\linewidth}
\begin{tabbing}\hspace{0.4cm}\=\hspace{0.8cm}\= \hspace{0.4cm}\= \\
\ {\bf procedure} {\sc LexFib}($pos$, $delta$: integer)\\
\>    {\bf if} ($pos=n+1$) {\bf print} $w$;\\
\>    {\bf else}\>     $w[pos]:=0$; \\
\>\>     {\bf if} ($w[pos-1]=1$) $d:=q-1$; {\bf else} $d$:=$delta+q$; {\bf endif} \ \\
\>\>     {\sc LexFib}($pos+1$, $d$);\\
\>\>      {\bf if} ($delta>0$)\\
\>\>\>       $w[pos]:=1$; {\sc LexFib}($pos+1$, $delta-1$);\\
\>\>      {\bf endif}\\
\>    {\bf endif}\\
\ {\bf end procedure}\\
\end{tabbing}
\end{minipage}
\\
\hline
\end{tabular}
\caption{\label{gen_alg} Lexicographic generation algorithm for $q$-decreasing words.}
\end{center}
\end{figure}
\comm{
Consider the image of $\Bw_2(00)$ under $\phi$:
$$ \phi(10) = 10, \; \phi(11) = 00, \; \phi(01) = 11. $$
We see
immediately, that $\phi$ does not preserve the Gray order, we can use
neither already known (for example~\cite{squire, vaj-fib}) nor any
other Gray code for classical Fibonacci words.}
\begin{table}[ht]
\begin{center}
{
\begin{tabular}{cc}
\begin{tabular}{c|c }
$u\in\Bw_4(111)$ & $\phi(u)\in\W^2_4$\\ \hline 
1100 & 0011\\ 
1101 & 1111\\ 
1001 & 1001\\ 
1000 & 0001\\ 
1010 & 0101\\ 
1011 & 1101\\ 
0011 & 1100\\
0010 & 0100\\ 
0000 & 0000\\ 
0001 & 1000\\ 
0101 & 1010\\ 
0100 & 0010\\ 
0110 & 1110\\
\end{tabular}
& \hspace{0.5cm}
\begin{tabular}{c c c|c c c}
\multicolumn{5}{c}{Words in $\W^1_6$ in BRGC order}    \\ \hline
{\it 1}  & 000000 &    &  {\it 12}  & 111100  &   3  \\
{\it 2}  & 000001 & 1  &  {\it 13}  & 111111  &   2 \\
{\it 3}  & 000011 & 1  &  {\it 14}  & 111110  &   1 \\
{\it 4}  & 000010 & 1  &  {\it 15}  & 111001  &   3 \\
{\it 5}  & 000110 & 1  &  {\it 16}  & 111000  &   1 \\
{\it 6}  & 000100 & 1  &  {\it 17}  & 100100  &   3 \\
{\it 7}  & 001001 & 3  &  {\it 18}  & 100010  &   2\\
{\it 8}  & 001000 & 1  &  {\it 19}  & 100011  &   1\\
{\it 9}  & 110000 & 3  &  {\it 20}  & 100001  &   1\\
{\it 10} & 110001 & 1  &  {\it 21}  & 100000  &   1\\
{\it 11} & 110010 & 2  &            &    &     \\
\end{tabular}\\
(a) & \hspace{0.5cm}(b)
\end{tabular}
}
\end{center}
\caption{\label{example_Gray}
(a) The images of words in $\Bw_4 \left(111\right)$ under the bijection $\phi$. Words in $\Bw_4\left(111\right)$ are listed in a BRGC-like order, called local reflected order in~\cite{vaj-fib}, which yields a 1-Gray code order. (b) The set $\W^1_6$ in BRGC order together with the Hamming distance between consecutive words.}
\end{table}

Generally, the bijection $\phi$ in relation (\ref{def-phi}) does not preserve 
{\it Graycodeness},
i.e. a Gray code for $(q+1)$-Fibonacci words is not necessarily
 mapped by $\phi$ to Gray code for $q$-decreasing words.
For instance, when $q=1$ and $n=2k+1$ the Gray
code for Fibonacci words in~\cite{vaj-fib} always contains two
consecutive words $u = (10)^k1$ and $v= (10)^k0$, but their images
$\phi(u) = 1^{2k+1}$ and $\phi(v) = 0^{k+1} 1^k$ have arbitrarily
large Hamming distance for enough large $n$.  A similar phenomenon
happens when $n=2k$ and $q=1$ with $u = (10)^{k-2}10$ and $v=
(10)^{k-2}00$: $\phi(u) = 1^{2k}$ and $\phi(v) = 0^{k+1}1^{k-1}$. See
also Table \ref{example_Gray}(a) for the image through $\phi$ of the
1-Gray code in \cite{vaj-fib} for $\Bw_4(111)$. 

Below we show that BRGC order (that is, the order induced by Binary
Reflected Gray Code in \cite{gray}) yields a 3-Gray code on $\W^q_n$.
Much more interestingly, thanks to Corollary \ref{parity}, the
  necessary condition for the existence of a $1$-Gray code is satisfied, and
  we  will provide such a Gray code for $\W^1_n$ in the following part.

\medskip

\noindent In~\cite{lin} the author introduces the notion of {\em
  absorbent set}, which (up to complement) is defined as: a binary
word set $\mathcal{X} \subset \{0,1\}^n$ is called {\rm absorbent} if
for any $u \in \mathcal{X}$ and any $k,$ $1 \leq k < n$, $u_1 u_2
\ldots u_k 0^{n-k}$ is also a word in $\mathcal{X}$.  Corollary 1 from
the same paper proves that when an absorbent set $\mathcal{X}$ is
listed in   BRGC  order (that is, restricting BRGC to $\mathcal{X}$) the resulting listing 
yields a 3-Gray code. Clearly, $\W^q_n$ is an absorbent set and we
have the following consequence.

\begin{corollary}
  For $n \ge 0, q \ge 1$, the restriction of BRGC order yields a
  3-Gray code for $\W^q_n$.
\label{vincentGray}
\end{corollary}

\noindent
See for an example Table \ref{example_Gray}(b).

Applying reversing sublists technique~\cite{Rus_93} by 
adding a new parameter to {\sc LexFib} which keeps track of the parity of the number of $1$s in the current generated prefix, procedure {\sc LexFib} can be turned into one generating the same class of words in BRGC order instead of the lexicographic one, and so according to Corollary \ref{vincentGray} producing a 3-Gray code for $\W^q_n$. 
Obviously the obtained Gray code generating algorithm inherits CAT property.

At this point it is worth mentioning an alternative powerful framework to design Gray codes as sublits of BRGC.
In \cite{SWW} it is given the following definition (up to mirroring each word): a set of same length binary words is a 
{\it flip-swap language} (with respect to $1$) if it is closed under the operations (i) flip the rightmost 1, and (ii) swap the rightmost $1$ with the bit to its left. And a flip-swap language with respect to $0$ is defined similarly.
Theorem 2 in \cite{SWW} states that when a flip-swap language is listed in BRGC order the resulting listing yields a 2-Gray code, and in the same paper there is given a variety of combinatorial classes that are flip-swap languages together with efficient generating 
algorithms for most of them.
It is easy to see that absorbent sets coincide with languages closed under 
flip operation, and thus it is not surprising that, when listed in BRGC order,
flip-swap languages yield more restrictive (that is, 2- instead of 3-) Gray codes than 
absorbent sets do. In particular, for $n \ge 0$ and $q \ge 1$, $\W^q_n$ is an absorbent 
set (or equivalently, a flip-closed language) but not a flip-swap language.

\subsection{$1$-Gray code for $\W^1_n$}

In this section, we construct a $1$-Gray code for the set $\W^q_n$,
$n\geq 0$, when $q=1$, which in particular gives a positive answer to
a conjecture in \cite{ege1}.  For this purpose, we decompose $\W^1_n$
$n\geq 1$, as $$\mathcal{W}^1_n=\mathcal{Z}_n \cup 1\cdot
\mathcal{W}^1_{n-1}$$ where $\W^1_0$
consists of the empty word
$\epsilon$, and $\mathcal{Z}_n$ is the subset of words starting with
$0$ in $\mathcal{W}^1_n$. In turn, we decompose $\mathcal{Z}_n$ as
$$\mathcal{Z}_n=\{0^n\}\cup\bigcup_{r=3}^{n} \mathcal{D}_n^r$$ where $\mathcal{D}_n^r=\bigcup_{j=1}^{\lfloor\frac{r-1}{2}\rfloor} 0^{r-j}1^{j} \cdot\mathcal{Z}_{n-r}$.
We refer to Figure~\ref{fig-big}(a) for a graphical illustration of
the decomposition of $\mathcal{Z}_{n}$ for $n=17$ where the point at coordinates
$(i,j)$ corresponds to the set $0^i1^j\cdot\mathcal{Z}_{n-i-j}$,
$1\leq j< i\leq n-1$, $3\leq i+j\leq n$, except the lowest point which
corresponds to $\{0^n\}$. The sets $\mathcal{D}^r_n$, $3\leq r\leq n$,
correspond to the southwest-northeast diagonals of the graphic.

\input{fig-big}
Let $\bb{Z}_0 = \epsilon$ be the list containing only the
  empty word $\epsilon$, $\bb{Z}_1 = 0, \bb{Z}_2 = 00, \bb{Z}_3 = 000
  \circ 001$.  
  If $\bb{L}$ is a list and $w$ is a word, then
  $w\cdot\bb{L}$ denotes the list where $w$ is concatenated to every
  word from $\bb{L}$. 
According to the above definitions, it is straightforward to check the
following lemma.

\begin{lemma}
  For any $1 \le k < n$, we suppose that $\bb{Z}_k$ is a $1$-Gray code
  for $\mathcal{Z}_k$ with $\first(\bb{Z}_k)=0(001)^\star$ and
  $\last(\bb{Z}_k)=(001)^\star$. Given $i$ and $j$ such that $1\leq
  j<i\leq n$, then
\begin{enumerate} 
\item[($i$)]
  for $3 \leq i+j<n$, the list
  $\bb{L}=0^i1^j \cdot \bb{Z}_{n-i-j}$ is a $1$-Gray code for $0^i1^j
  \cdot \mathcal{Z}_{n-i-j}$ with $\first(\bb{L})=0^i1^j0(001)^\star$
  and $\last(\bb{L})=0^i1^j(001)^\star$;\\
  when $i+j = n$ the list $\bb{L}$ contains only the word $0^i1^j$;
\item[($ii$)]
  for $3 \leq i + j < n-1$,
  the list
  $\bb{L}=0^i1^{j+1} \cdot \bb{Z}_{n-i-j-1}\circ 0^i1^{j} \cdot
  \bb{Z}_{n-i-j} $ is a $1$-Gray code for
  $
  0^i1^{j+1} \cdot \mathcal{Z}_{n-i-j-1}
  \cup 
  0^i1^j \cdot \mathcal{Z}_{n-i-j}
  $ with
  $\first( \bb{L})=0^i1^{j+1}0(001)^\star$ and
  $\last(\bb{L})=0^i1^{j}(001)^\star$;\\
  when $i+j +1 = n$, the list 
  $\bb{L} = 0^i1^{j+1} \circ 0^i1^j0$;
\item[($iii$)]
  for $3 \leq i + j < n$,
  the list
  $\bb{L}=0^i1^{j} \cdot \bb{Z}_{n-i-j}\circ \reverse{0^{i-1}1^{j+1}
  \cdot \bb{Z}_{n-i-j}} $ is a $1$-Gray code for $0^i1^j \cdot
  \mathcal{Z}_{n-i-j}\cup 0^{i-1}1^{j+1} \cdot \mathcal{Z}_{n-i-j}$
  with $\first( \bb{L})=0^i1^{j}0(001)^\star$ and
  $\last(\bb{L})=0^{i-1}1^{j+1}0(001)^\star$;\\
  when $i + j = n$, the list 
  $\bb{L} = 0^i1^j \circ 0^{i-1}1^{j+1}$;
\item[($iv$)]
  for $3 \leq i + j \le n$,
  the list
  $\bb{L}=\reverse{0^i1^{j} \cdot \bb{Z}_{n-i-j}}\circ 0^{i-1}1^{j+1}
  \cdot \bb{Z}_{n-i-j}$ is a $1$-Gray code for
  $0^i1^j \cdot
  \mathcal{Z}_{n-i-j}\cup 0^{i-1}1^{j+1} \cdot \mathcal{Z}_{n-i-j}$
  with $\first( \bb{L})=0^i1^{j}(001)^\star$ and
  $\last(\bb{L})=0^{i-1}1^{j+1}(001)^\star$.
\end{enumerate}
\label{lemma2}
\end{lemma}

\begin{lemma}
  Let us consider $r=1\mod{4}$, $3\leq r\leq n$. For any $1 \le k < n$, we
  suppose that $\bb{Z}_k$ is a $1$-Gray code for $\mathcal{Z}_k$ with
  $\first(\bb{Z}_k)=0(001)^\star$ and $\last(\bb{Z}_k)=(001)^\star$.
\begin{enumerate}

\item[($i$)] If $r\neq n-1$, then there is a $1$-Gray code $\Delta^r_n$
  for $\mathcal{D}^r_n\cup \mathcal{D}^{r-1}_n$ such that
  $\first(\Delta^r_n)=0^{r-2}1(001)^\star$ and
  $\last(\Delta^r_n)=0^{r-1}1(001)^\star$.
  
\item[($ii$)] If $r=n-1$, then there is a $1$-Gray code
  $\Delta^{n-1}_n$ for $\mathcal{D}^{n}_n\cup \mathcal{D}^{n-1}_n\cup
  \mathcal{D}^{n-2}_n$ such that $\first(\Delta^{n-1}_n)=0^{n-3}100$
  and $\last(\Delta^{n-1}_n)=0^{n-1}1$.
\end{enumerate}
\label{lemma3}
\end{lemma}

\proof
For the first assertion ($i$), it suffices to consider the list
  $$ \Delta_n^r=
  \bigcirc_{j=1}^{\frac{r-3}{2}}0^{r-1-j}1^j\cdot\langle\bb{Z}_{n-r+1}\rangle^{j}
  \circ\reverse{\bigcirc_{j=1}^{\frac{r-1}{2}}0^{r-j}1^j\cdot
    \langle\bb{Z}_{n-r}\rangle^{j}}
  .$$

After considering assertions of Lemma~\ref{lemma2}, it remains to examine the transition between
$w_0=\last (0^{r-1-j_0}1^{j_0}\cdot\langle\bb{Z}_{n-r+1}\rangle^{j_0})$
for $j_0=\frac{r-3}{2}$ and
$w_1= \first(\reverse{
0^{r-j_1}1^{j_1}\cdot\langle\bb{Z}_{n-r}\rangle^{j_1}})$
for $j_1=\frac{r-1}{2}$. Since $r=1\mod{4}$, we have necessarily
$j_0$ is odd and $j_0+1=j_1$ which implies that 
       $w_0 = \last (0^{r-1-j_0}1^{j_0}\cdot \reverse{\bb{Z}_{n-r+1}} )
       = 0^{r-1-j_0}1^{j_0} 0 (001)^\star $
       and
       $w_1 =
       \first(
       \reverse{
       0^{r-j_1}1^{j_1}\cdot\bb{Z}_{n-r}}
       )
       =
       0^{r-j_1}1^{j_1} (001)^\star $, and they
differ by exactly one
bit.

For the second assertion ($ii$), we consider the list
$$ \Delta_n^{n-1}=\bigcirc_{j=1}^{\frac{n-4}{2}}0^{n-2-j}1^j00\circ
0^{\frac{n}{2}}1^{\frac{n-2}{2}}0\circ
\reverse{
\bigcirc_{j=1}^{\frac{n-4}{2}}
\left\langle
  0^{n-j-1}1^{j+1}\circ 0^{n-1-j}1^j0
\right\rangle^{j-1}     
}
\circ 0^{n-1}1.$$

A simple study of each
kind of transitions allows us to see that $\Delta_n^{n-1}$ is a $1$-Gray
code for $\mathcal{D}^{n}_n\cup \mathcal{D}^{n-1}_n\cup
\mathcal{D}^{n-2}_n$ satisfying $\first(\Delta^{n-1}_n)=0^{n-3}100$
and $\last(\Delta^{n-1}_n)=0^{n-1}1$. An illustration of this Gray
code for $n=10$ (and thus $r=9$) can be found in the last sketch of
Figure \ref{fig-sequence}.\qed 


In the following, we write $w\bb{L}$ instead of $w\cdot\bb{L}$ to be more concise.

\begin{lemma}
  Let us consider $r=3\mod{4}$, $3<n$ and $3\leq r\leq n$. For any $1
  \le k < n$, we suppose that $\bb{Z}_k$ is a $1$-Gray code for
  $\mathcal{Z}_k$ with $\first(\bb{Z}_k)=0(001)^\star$ and
  $\last(\bb{Z}_k)=(001)^\star$.

\begin{enumerate}

\item[($i$)] If $r=3$, then there is a $1$-Gray code $\Delta^3_n$ for
  $\mathcal{D}^3_n$ such that $\first(\Delta^3_n)=0010(001)^\star$ and
  $\last(\Delta^3_n)=(001)^\star$.

\item[($ii$)] If $r=n-2$, then there is a $1$-Gray code
  $\Delta^{n-2}_n$ for $\mathcal{D}^{n-2}_n\cup \mathcal{D}^{n-3}_n$
  such that $\first(\Delta^{n-2}_n)=0^{n-4}1000$ and
  $\last(\Delta^{n-2}_n)=0^{n-3}100$.

\item[($iii$)] If $r=n-1$, then there is a $1$-Gray code
  $\Delta^{n-1}_n$ for $\mathcal{D}^{n}_n\cup\mathcal{D}^{n-1}_n\cup
  \mathcal{D}^{n-2}_n\setminus\{0^{n-1}1\}$ such that
  $\first(\Delta^{n-1}_n)=0^{n-3}100$ and
  $\last(\Delta^{n-1}_n)=0^{n-2}10$.

\item[($iv$)] If $r=n$, then there is a $1$-Gray code $\Delta^{n}_n$
  for $\mathcal{D}^{n}_n\cup \mathcal{D}^{n-1}_n$ such that
  $\first(\Delta^{n}_n)=0^{n-2}10$ and $\last(\Delta^{n}_n)=0^{n-1}1$.

\item[($v$)] If $r\notin\{3,n-2,n-1,n\}$, then there is a $1$-Gray code
  $\Delta^r_n$ for $\mathcal{D}^r_n\cup \mathcal{D}^{r-1}_n$ such that
  $\first(\Delta^r_n)=0^{r-1}10(001)^\star$ and
  $\last(\Delta^r_n)=0^{r-1}1(001)^\star$.

\end{enumerate}
\label{lemma4}
\end{lemma}
\proof
For the case ($i$),  we set: $\Delta_3=0^21\bb{Z}_{n-3}$.

For the case ($ii$), we set:
  $\Delta_n^{n-2}=\bigcirc_{j=1}^{\frac{n-5}{2}} 0^{n-3-j}1^j
  \langle\bb{Z}_{3}\rangle^{j-1}
  \circ \reverse{\bigcirc_{j=1}^{\frac{n-3}{2}} 0^{n-2-j}1^j\bb{Z}_{2}}$.
  Since we have $\bb{Z}_2=00$ and $\bb{Z}_3=000 \circ 001$, it is
  straightforward to see that $\Delta_n^{n-2}$ is a $1$-Gray code with
  $\first(\Delta_n^{n-2})=0^{n-4}1000$ and
  $\last(\Delta^{n-2}_n)=0^{n-3}100$.

  For the case ($iii$),  we set:
  $$\Delta_n^{n-1}=\bigcirc_{j=1}^{\frac{n-4}{2}} 0^{n-2-j}1^j\bb{Z}_2\circ 0^{\frac{n}{2}}1^{\frac{n-2}{2}}\bb{Z}_1 \circ
  \reverse{\bigcirc_{j=1}^{\frac{n-4}{2}}
    \left\langle0^{n-1-j}1^{j+1}\circ 0^{n-1-j}1^j\bb{Z}_{1}\right\rangle^{j}}.$$

Knowing that $\bb{Z}_2=00$ and $\bb{Z}_1=0$, we can easily check that any pair of consecutive words differ by exactly one bit, which proves that $\Delta_n^{n-1}$  is a $1$-Gray code.

 For the case ($iv$),  we set: $\Delta_n^{n}=\bigcirc_{j=1}^{\frac{n-3}{2}} 0^{n-1-j}1^{j}0\circ \reverse{\bigcirc_{j=1}^{\frac{n-1}{2}} 0^{n-j}1^j}.$
As previously the result can be obtained easily.

The case ($v$) is more challenging to handle. The set
$\mathcal{D}^r_n\cup \mathcal{D}^{r-1}_n$ consists of the union of the
following subsets: $K_1=0^{r-2}1\mathcal{Z}_{n-r+1}$,
$K_2=0^{r-3}11\mathcal{Z}_{n-r+1}$, $\ldots$,
$K_a=0^{r-a-1}1^{a}\mathcal{Z}_{n-r+1}$
and
$L_1=0^{r-1}1\mathcal{Z}_{n-r}$, $L_2=0^{r-2}11\mathcal{Z}_{n-r}$,
$\ldots$,
$L_{b}=0^{\frac{r+1}{2}}1^{\frac{r-1}{2}}\mathcal{Z}_{n-r}$
$L_{a+1}=0^{r-a-1}1^{a+1}\mathcal{Z}_{n-r}$
with $a=\lfloor\frac{r-2}{2}\rfloor=\frac{r-3}{2}$.
Let us denote by $\bb{K}_1,
\bb{K}_2, \ldots , \bb{K}_a$ and $\bb{L}_1, \bb{L}_2, \ldots ,\bb{L}_{a+1}$
the associated Gray codes obtained by replacing
$\mathcal{Z}_k$ with the Gray code $\bb{Z}_k$.

Remark that for $1\leq i\leq a-1$ (resp. $1\leq i\leq a$) and for a
given $j$, the $j$th word of $\bb{K}_i$ (resp. $\bb{L}_i$) and the
$j$th word of $\bb{K}_{i+1}$ (resp. $\bb{L}_{i+1}$) differ by exactly
one bit.
Recall that $\widehat{\bb{A}}$ is obtained from a list $\bb{A}$ by deleting its first element. The words $\first(\bb{K}_i)$ and $\last(\bb{L}_{i+1})$
differ by one bit. We have 
$\first(\widehat{\bb{L}_{a+1}} \circ \bb{K}_a) = \first (\widehat{\bb{L}_{a+1}})$
and it differs obviously by one bit from $\first(\bb{L}_{a+1})$.
We also have $r=3\mod{4}$, so $\frac{r-3}{2}=a$ is even.
Taking into account all these remarks,
the list $\Delta_n^r$ defined below is a Gray code:

$$
  \Delta_n^r =
  \bigcirc_{i=1}^{a+1} \first(\bb{L}_i) \circ \bigcirc_{i=a}^1
  \big\langle \widehat{\bb{L}_{i+1}} \circ \bb{K}_i \big\rangle^i
  \circ \widehat{\bb{L}_1}
$$
We refer to Figure~\ref{figaaas} for a graphical representation of
this Gray code.\qed
\begin{figure}[ht] 
	\begin{center}
	  \scalebox{1}{\begin{tikzpicture}[scale=1.8]
	      \tikzset{q/.style={dotted, gray, line width=4pt}};

        \draw[q] (0,0.3)--(0,1.45);
				\draw[q] (1,0.3)--(1,1.45);
				\draw[q] (1,2.25)--(1,3.35);
				\draw[q] (2,0.3)--(2,1.45);
				\draw[q] (2,2.25)--(2,3.35);
				\draw[q] (3,0.3)--(3,1.45);
				\draw[q] (3,2.25)--(3,3.35);
				\draw[q] (4,0.3)--(4,1.45);	
				\draw[q] (4,2.25)--(4,3.35);
				\draw[q] (5,0.3)--(5,1.45);
				\draw[q] (5,2.25)--(5,3.35);
				\draw[q] (6,0.3)--(6,1.45);	
				\draw[q] (6,2.25)--(6,3.35);
        \draw (-0.3,1) node {\scriptsize$\bb{L}_1$};
        \draw (1.3,1) node {\scriptsize$\bb{L}_2$};
        \draw (1.7,1) node {\scriptsize$\bb{L}_3$};
        \draw (3.3,1) node {\scriptsize$\bb{L}_4$};
        \draw (3.7,1) node {\scriptsize$\bb{L}_5$};
        \draw (5.3,1) node {\scriptsize$\bb{L}_6$};
        \draw (5.7,1) node {\scriptsize$\bb{L}_7$};
        \draw (1.3,2.9) node {\scriptsize$\bb{K}_1$};
        \draw (1.7,2.9) node {\scriptsize$\bb{K}_2$};
        \draw (3.3,2.9) node {\scriptsize$\bb{K}_3$};
        \draw (3.7,2.9) node {\scriptsize$\bb{K}_4$};
        \draw (5.3,2.9) node {\scriptsize$\bb{K}_5$};
        \draw (5.7,2.9) node {\scriptsize$\bb{K}_6$};

        \tikzset{p/.style={radius=1.5pt}};
        \fill (0,0.05) circle[p];
        \fill (0,0.2) circle[p];
        \fill (0,1.4) circle[p];

        \fill (1,0.05) circle[p];
        \fill (1,0.2) circle[p];
        \fill (1,1.4) circle[p];
        \fill (1,2.15) circle[p];
        \fill (1,3.44) circle[p];
        
        \fill (2,0.05) circle[p];
        \fill (2,0.2) circle[p];
        \fill (2,1.4) circle[p];
        \fill (2,2.15) circle[p];
        \fill (2,3.44) circle[p];

        \fill (3,0.05) circle[p];
        \fill (3,0.2) circle[p];
        \fill (3,1.4) circle[p];
        \fill (3,2.15) circle[p];
        \fill (3,3.44) circle[p];

        \fill (4,0.05) circle[p];
        \fill (4,0.2) circle[p];
        \fill (4,1.4) circle[p];
        \fill (4,2.15) circle[p];
        \fill (4,3.44) circle[p];

        \fill (5,0.05) circle[p];
        \fill (5,0.2) circle[p];
        \fill (5,1.4) circle[p];
        \fill (5,2.15) circle[p];
        \fill (5,3.44) circle[p];

        \fill (6,0.05) circle[p];
        \fill (6,0.2) circle[p];
        \fill (6,1.4) circle[p];
        \fill (6,2.15) circle[p];
        \fill (6,3.44) circle[p];

        \draw [black, line width=1.7pt,
          decoration={markings,
            mark=at position 0.999 with
            {
              \arrow[scale=1.2,]{latex}
            }
          },
          shorten >= 2pt,
          postaction={decorate}
        ] (-0.02, 0.05) -- 
        (6, 0.05) -- 
        (6, 3.44) --
        (5, 3.44) -- (5, 0.2) -- (4, 0.2) -- 
        (4, 3.44) --
        (3, 3.44) -- (3, 0.2) -- (2, 0.2) -- 
        (2, 3.44) --
        (1, 3.44) -- (1, 0.2) -- (0, 0.2) -- 
        (0, 1.4)
        ;

        \tikzset{l/.style={fill=white}};
        
        \draw (0,-0.1) node {\tiny$0^{r-1}10(001)^\star$};
        \draw (1,-0.1) node {\tiny$0^{r-2}1^20(001)^\star$};
        \draw (2,-0.1) node {\tiny$0^{r-3}1^30(001)^\star$};
        \draw (3,-0.1) node {\tiny$0^{r-4}1^40(001)^\star$};
        \draw (4,-0.1) node {\tiny$0^{r-5}1^50(001)^\star$};
        \draw (5,-0.1) node {\tiny$0^{r-6}1^60(001)^\star$};        
        \draw (6,-0.1) node {\tiny$0^{r-7}1^70(001)^\star$};

        \draw (0,1.6) node {\tiny$0^{r-1}1(001)^\star$};
        \draw (1,1.6) node[l] {\tiny$0^{r-2}1^2(001)^\star$};
        \draw (2,1.6) node[l] {\tiny$0^{r-3}1^3(001)^\star$};
        \draw (3,1.6) node[l] {\tiny$0^{r-4}1^4(001)^\star$};
        \draw (4,1.6) node[l] {\tiny$0^{r-5}1^5(001)^\star$};
        \draw (5,1.6) node[l] {\tiny$0^{r-6}1^6(001)^\star$};        
        \draw (6,1.6) node[l] {\tiny$0^{r-7}1^7(001)^\star$};

        \draw (1,1.95) node[l] {\tiny$0^{r-2}10(001)^\star$};
        \draw (2,1.95) node[l] {\tiny$0^{r-3}1^20(001)^\star$};
        \draw (3,1.95) node[l] {\tiny$0^{r-4}1^30(001)^\star$};
        \draw (4,1.95) node[l] {\tiny$0^{r-5}1^40(001)^\star$};
        \draw (5,1.95) node[l] {\tiny$0^{r-6}1^50(001)^\star$};        
        \draw (6,1.95) node[l] {\tiny$0^{r-7}1^60(001)^\star$};

        \draw (1,3.65) node[l] {\tiny$0^{r-2}1(001)^\star$};
        \draw (2,3.65) node[l] {\tiny$0^{r-3}1^2(001)^\star$};
        \draw (3,3.65) node[l] {\tiny$0^{r-4}1^3(001)^\star$};
        \draw (4,3.65) node[l] {\tiny$0^{r-5}1^4(001)^\star$};
        \draw (5,3.65) node[l] {\tiny$0^{r-6}1^5(001)^\star$};        
        \draw (6,3.65) node[l] {\tiny$0^{r-7}1^6(001)^\star$};

      \end{tikzpicture}
    }
	\end{center}
	\caption{ An illustration of the Gray code $\Delta_n^r$ for the case
    ($v$) in the proof of Lemma 4 (we consider $a=6$, $r= 15$). Vertical
    sequences of squares are Gray codes $\bb{K}_i$, $1\leq i\leq a$,
    and $\bb{L}_i$, $1\leq i\leq a+1$, so that the first and the last
    elements are respectively the bottom and top squares of the
    segments. The walk illustrates the Gray code $\Delta_n^r$ that
    starts with $\first(\bb{L}_1)$ and ends with $\last(\bb{L}_1)$.}
\label{figaaas}
\end{figure}
\endproof

\begin{theorem}
  For any $n \ge 0$, there exists a $1$-Gray code $\bb{Z}_n$ for
  $\mathcal{Z}_n$ such that $\first(\bb{Z}_n)=0(001)^\star$ and
  $\last(\bb{Z}_n)=(001)^\star$.
  \label{Z}
\end{theorem}

Using initial conditions $\bb{Z}_0 = \epsilon,
  \bb{Z}_1 = 0, \bb{Z}_2 = 00, \bb{Z}_3 = 000 \circ 001$, where
  $\epsilon$ is the empty word, and the recursive constructions for
  lists $\Delta^r_n$ (Lemmas~\ref{lemma2} and~\ref{lemma3}) we
  define the $1$-Gray code $\bb{Z}_n$ as follows:
  
  \vspace{1ex}
  \centerline{\scalebox{0.87}{$\bb{Z}_n = \begin{cases} 
     \Delta_n^5 \circ \Delta_n^9 \circ \cdots
    \circ \Delta_n^{n}\circ 0^n \circ\Delta_n^{n-2}\circ
      \cdots \circ \Delta_n^7 \circ \Delta_n^3
    &
    \text{ if } n = 1\mod 4, \\
      \Delta_n^5 \circ \Delta_n^9 \circ \cdots
    \circ \Delta_n^{n-1}\circ 0^n \circ\Delta_n^{n-3}\circ
      \cdots \circ \Delta_n^7 \circ \Delta_n^3
    &
    \text{ if } n =2\mod 4, \\
      \Delta_n^5 \circ \Delta_n^9 \circ \cdots
    \circ \Delta_n^{n-2}\circ 0^n \circ\Delta_n^{n}\circ
      \cdots \circ \Delta_n^7 \circ \Delta_n^3
    &
    \text{ if } n = 3\mod 4. \\
    \Delta_n^5 \circ \Delta_n^9 \circ \cdots \circ\Delta_n^{n-3}\circ 0^{n-1}1 \circ 0^n \circ\Delta_n^{n-1} \circ
    \cdots \circ \Delta_n^7 \circ \Delta_n^3
    &\text{ if } n= 0 \mod 4.\\
  \end{cases}$}}


 Due to Lemmas~\ref{lemma3} and~\ref{lemma4},
   $\last(\Delta_n^{4i+1})$ differs by one bit from
   $\first(\Delta_n^{4i+5})$ and $\first(\Delta_n^{4i+3})$ differs by
   one bit from $\last(\Delta_n^{4i+7})$. Using $0^n$ (and $0^{n-1}1$
   when $n = 0 \mod 4$) we connected all $\Delta^r_n$ and ensure that
   $\bb{Z}_n$ is a $1$-Gray code. See Table~\ref{teltas} for a more detailed view on the structure of $\bb{Z}_n$.
 \qed
 \endproof

 \begin{table}[ht]
   \scriptsize
   \begin{tabular}{|r|l|l|}
     \multicolumn{3}{c}{$n = 1 \mod 4$}       \\
     \hline
     \parbox[t]{4mm}{
       \multirow{13}{*}{
       \rotatebox[origin=c]{90}{Lemma~\ref{lemma3}}
       }
     }
     & \multirow{3}{*}{$\Delta^5_n =$}  & $0001(001)^\star$     \\
     & & $\cdots$             \\
     & & $00001(001)^\star$    \\
     \cline{2-3}
     & \multirow{3}{*}{$\Delta^9_n =$}  & $00000\;001(001)^\star$     \\
     & & $\cdots$             \\
     & & $000000001(001)^\star$    \\
     \cline{2-3}
     & $\cdots$ & $\cdots$ \\
     \cline{2-3}
     & \multirow{3}{*}{$\Delta^{n-4}_n =$}  & $0^{n-6}1(001)^\star$     \\
     & & $\cdots$             \\
     & & $0^{n-5}1(001)^\star$    \\
     \cline{2-3}     
     & \multirow{3}{*}{$\Delta^n_n =$}  & $0^{n-2}10$     \\
     & & $\cdots$             \\
     & & $0^{n-1}1$    \\
     \hline
     & &  $0^n$ \\
     \hline
     \parbox[t]{4mm}{
       \multirow{13}{*}{
         \rotatebox[origin=c]{90}{Lemma~\ref{lemma4}}
       }
     }
     & \multirow{3}{*}{$\Delta^{n-2}_n =$}  & $0^{n-4}1000$     \\
     &  & $\cdots$     \\
     &  & $0^{n-3}100$     \\
     \cline{2-3}
     & \multirow{3}{*}{$\Delta^{n-6}_n =$}  & $0^{n-7}10(001)^\star$     \\
     &  & $\cdots$     \\
     &  & $0^{n-7}1(001)^\star$     \\
     \cline{2-3}
     & $\cdots$ & $\cdots$ \\
     \cline{2-3}     
     & \multirow{3}{*}{$\Delta^7_n =$}  & $00000010(001)^\star$     \\
     & & $\cdots$             \\
     & & $0000\;001(001)^\star$    \\
     \cline{2-3}
     & \multirow{3}{*}{$\Delta^3_n =$}  & $0010(001)^\star$     \\
     & & $\cdots$             \\
     & & $(001)^\star$    \\
     \hline
   \end{tabular}\,%
   \begin{tabular}{|l|l|}
      \multicolumn{2}{c}{$n = 2 \mod 4$}       \\
     \hline
      \multirow{3}{*}{$\Delta^5_n =$}  & $0001(001)^\star$     \\
      & $\cdots$             \\
      & $00001(001)^\star$    \\
     \hline
      \multirow{3}{*}{$\Delta^9_n =$}  & $00000\;001(001)^\star$     \\
      & $\cdots$             \\
      & $000000001(001)^\star$    \\
     \hline
      $\cdots$ & $\cdots$ \\
     \hline
      \multirow{3}{*}{$\Delta^{n-5}_n =$}  & $0^{n-7}1(001)^\star$     \\
      & $\cdots$             \\
      & $0^{n-6}1(001)^\star$    \\
     \hline
      \multirow{3}{*}{$\Delta^{n-1}_n =$}  & $0^{n-3}100$     \\
      & $\cdots$             \\
      & $0^{n-1}1$    \\
     \hline
      &  $0^n$ \\
     \hline
     \multirow{3}{*}{$\Delta^{n-3}_n =$}  & $0^{n-4}10(001)^\star$     \\
       & $\cdots$     \\
       & $0^{n-4}1(001)^\star$     \\
     \hline
      \multirow{3}{*}{$\Delta^{n-7}_n =$}  & $0^{n-8}10(001)^\star$     \\
       & $\cdots$     \\
       & $0^{n-8}1(001)^\star$     \\
     \hline
      $\cdots$ & $\cdots$ \\
     \hline
      \multirow{3}{*}{$\Delta^7_n =$}  & $00000010(001)^\star$     \\
      & $\cdots$             \\
      & $0000\;001(001)^\star$    \\
     \hline
     \multirow{3}{*}{$\Delta^3_n =$}  & $0010(001)^\star$     \\
      & $\cdots$             \\
      & $(001)^\star$    \\
     \hline
   \end{tabular}\,%
   \begin{tabular}{|l|l|}
     \multicolumn{2}{c}{$n = 3 \mod 4$}       \\
     \hline
     \multirow{3}{*}{$\Delta^5_n =$}  & $0001(001)^\star$     \\
     & $\cdots$             \\
     & $00001(001)^\star$    \\
     \hline
     \multirow{3}{*}{$\Delta^9_n =$}  & $00000\;001(001)^\star$     \\
     & $\cdots$             \\
     & $000000001(001)^\star$    \\
     \hline
     $\cdots$ & $\cdots$ \\
     \hline
     \multirow{3}{*}{$\Delta^{n-6}_n =$}  & $0^{n-8}1(001)^\star$     \\
     & $\cdots$             \\
     & $0^{n-7}1(001)^\star$    \\
     \hline
     \multirow{3}{*}{$\Delta^{n-2}_n =$}  & $0^{n-4}1(001)^\star$     \\
     & $\cdots$             \\
     & $0^{n-3}1(001)^\star$    \\
     \hline
     &  $0^n$ \\
     \hline
     \multirow{3}{*}{$\Delta^{n}_n =$}  & $0^{n-2}10$     \\
     & $\cdots$     \\
     & $0^{n-1}1$     \\
     \hline
     \multirow{3}{*}{$\Delta^{n-4}_n =$}  & $0^{n-5}10(001)^\star$     \\
     & $\cdots$     \\
     & $0^{n-5}1(001)^\star$     \\
     \hline
     $\cdots$ & $\cdots$ \\
     \hline
     \multirow{3}{*}{$\Delta^7_n =$}  & $00000010(001)^\star$     \\
     & $\cdots$             \\
     & $0000\;001(001)^\star$    \\
     \hline
     \multirow{3}{*}{$\Delta^3_n =$}  & $0010(001)^\star$     \\
     & $\cdots$             \\
     & $(001)^\star$    \\
     \hline
   \end{tabular}\,%
   \begin{tabular}{|l|l|}
     \multicolumn{2}{c}{$n = 0 \mod 4$}       \\
     \hline
     \multirow{3}{*}{$\Delta^5_n =$}  & $0001(001)^\star$     \\
     & $\cdots$             \\
     & $00001(001)^\star$    \\
     \hline
     \multirow{3}{*}{$\Delta^9_n =$}  & $00000\;001(001)^\star$     \\
     & $\cdots$             \\
     & $000000001(001)^\star$    \\
     \hline
     $\cdots$ & $\cdots$ \\
     \hline
     \multirow{3}{*}{$\Delta^{n-7}_n =$}  & $0^{n-9}1(001)^\star$     \\
     & $\cdots$             \\
     & $0^{n-8}1(001)^\star$    \\
     \hline
     \multirow{3}{*}{$\Delta^{n-3}_n =$}  & $0^{n-5}1(001)^\star$     \\
     & $\cdots$             \\
     & $0^{n-4}1(001)^\star$    \\
     \hline
     &  $0^{n-1}1$ \\
     &  $0^n$ \\
     \hline
     \multirow{3}{*}{$\Delta^{n-1}_n =$}  & $0^{n-3}100$     \\
     & $\cdots$     \\
     & $0^{n-2}10$     \\
     \hline
     \multirow{3}{*}{$\Delta^{n-5}_n =$}  & $0^{n-6}10(001)^\star$     \\
     & $\cdots$     \\
     & $0^{n-6}1(001)^\star$     \\
     \hline
     $\cdots$ & $\cdots$ \\
     \hline
     \multirow{3}{*}{$\Delta^7_n =$}  & $00000010(001)^\star$     \\
     & $\cdots$             \\
     & $0000\;001(001)^\star$    \\
     \hline
     \multirow{3}{*}{$\Delta^3_n =$}  & $0010(001)^\star$     \\
     & $\cdots$             \\
     & $(001)^\star$    \\
     \hline
   \end{tabular}
   
   \caption{The structure of $\bb{Z}_n$.}
   \label{teltas}
\end{table}

We refer to Figure~\ref{fig-sequence} for a graphical representation
of $\bb{Z}_n$ for $4\leq n\leq 10$, see also Figure~\ref{fig-big}(b)
for $n=17$. An immediate consequence of Theorem~\ref{Z} is the
following.

\begin{theorem} For any $n\geq 1$, $\bb{W}^1_n=1 \cdot \bb{W}^1_{n-1} \circ \bb{Z}_n$ is a $1$-Gray code  for $\mathcal{W}^1_n$ such that
  $\first(\bb{W}^1_n)=1^n$, $\last(\bb{W}^1_n)=(001)^\star$, and where
  $\bb{W}^1_0$ is a list containing only the empty word.
\label{Th4}
\end{theorem}

However, the efficient generation of this Gray code remains an
   open problem.

\begin{table}[ht]
\centering
\begin{tabular}{c p{1.3cm} |c p{1.3cm} |c c }
\hline
{\it 1}  & 111111 &     {\it 8}  & 110010  & {\it 15} &  000110 \\
{\it 2}  & 111110 &    {\it 9}  & 100010    & {\it 16} & 000010\\
{\it 3}  & 111100 &   {\it 10}  &  100011  & {\it 17} & 000011\\
{\it 4}  & 111000 &   {\it 11}  & 100001 & {\it 18} & 000001 \\
{\it 5}  & 111001 &   {\it 12}  &  100000  &  {\it 19}&000000 \\
{\it 6}  & 110001 &   {\it 13}  & 100100 &  {\it 20}&001000\\
{\it 7}  & 110000 &   {\it 14}  & 000100 &  {\it 21}& 001001\\
\end{tabular}
\caption{The Gray code $\bb{W}^1_6$ for the set $\mathcal{W}^1_6$. The Hamming distance between two consecutive words is one.}
\label{example_Graycode}
\end{table}

Eğecioğlu and Iršič introduce in~\cite{ege1} the {\em
    ``run-constrained binary strings''}. These are binary words, in
  which every run of 1s is immediately followed by a strictly longer
  run of 0s.  Using these strings of length $n+2$ as vertices, and
  connecting two vertices if they differ at only one position, the authors
  of~\cite{ege1} form the {\em Fibonacci-run graph} $\mathcal{R}_n$ as
  the induced subgraph of the hypercube. (As every non-empty
  run-constrained string must end with 00, authors of~\cite{ege1}
  actually drop the last 2 zeros, but not we.)  Figure~\ref{fibograph}
  gives small examples.

\input{fibograph}

It turns out that the run-constrained binary strings are precisely the
reverse of $1$-decreasing words beginning with $0$.  In this light, the
Gray code $\bb{Z}_n$ in Theorem~\ref{Z} gives a Hamiltonian path in
the Fibonacci-run graph.  The next corollary settles a conjecture
in~\cite{ege1}.

\begin{corollary}
  For any $n \ge 1$, the Fibonacci-run graph $\mathcal{R}_n$ has a
  Hamiltonian path.
  \label{conj-solved}
\end{corollary}

\noindent Lemma 9.1 from~\cite{ege1} says that if $n \ne 1 \mod 3$,
then $\mathcal{R}_n$ does not contain a Hamiltonian cycle.  Our method
give a Hamiltonian path, which is not a cycle.  The question of
whether there is a Hamiltonian cycle for the case $n = 1 \mod 3$
remains open. 

Finally, the validity of the parity condition stated in Corollary~\ref{parity}
and experimental investigations for small values, $0 \le n \le 5$ and
$2 \le q \le 5$, suggest the following extension of Theorem \ref{Th4}.

\begin{conj} For any $n \ge 0$ and $q\geq 1$, there is a $1$-Gray code for $\mathcal{W}^q_n$.
\end{conj}

\input{fig-sequence}


\noindent
{\bf Acknowledgment.}
The authors are grateful to the anonymous referees for helping to rectify an error in the construction of the $1$-Gray code in the last part of the paper and 
for providing numerous comments helping to improve the presentation of the paper.

\bibliographystyle{splncs04}
\bibliography{bib}

\end{document}

%% file: fig-big.tex
\begin{figure}[!ht]
    \begin{subfigure}[b]{0.38\textwidth}
         \centering
         \input{fig-big-rien.tex}
         \caption{}
         \label{fig-big:a}
    \end{subfigure}
     \begin{subfigure}[b]{0.5\textwidth}
       \centering
         \tikzset{
    b/.style={circle, fill, minimum size=1mm, inner sep=0pt, draw=white},
    sp/.style={line width=1mm, white},
    every label/.style={label distance=2mm}
  }
  \tt
  \scriptsize
  \begin{tikzpicture}[scale=0.5]
    \node[b,label=left:\tt001..............] (a001) at (0,0) {};

    \node[b,label=left:\tt0001.............] (a0001) at (0,-1) {};
    \node[b,label={[label distance=1mm]right:00011............}] (a00011) at (1,-1) {};

    \node[b,label=left:00001............] (a00001) at (0,-2) {};
    \node[b] (a000011) at (1,-2) {};
    \node[b,label={[label distance=1mm]right:0000111..........}] (a0000111) at (2,-2) {};

    \node[b,label=left:{000001...........}] (a000001) at (0,-3) {};
    \node[b] (a0000011) at (1,-3) {};
    \node[b] (a00000111) at (2,-3) {};
    \node[b,label={[label distance=1mm]right:000001111........}] (a000001111) at (3,-3) {};

    \node[b,label=left: {0000001..........}] (a0000001) at (0,-4) {};
    \node[b] (a00000011) at (1,-4) {};
    \node[b] (a000000111) at (2,-4) {};
    \node[b] (a0000001111) at (3,-4) {};
    \node[b,label=right:{00000011111......}] (a00000011111) at (4,-4) {};
    
    \node[b,label=left:{$\tt 0^71.........$}] (a071) at (0,-5) {};
    \node[b] (a0711) at (1,-5) {};
    \node[b] (a07111) at (2,-5) {};
    \node[b] (a071111) at (3,-5) {};
    \node[b] (a0711111) at (4,-5) {};
    \node[b,label=right:{$\tt 0^7111111....$}] (a07111111) at (5,-5) {};
    
    \node[b,label=left:{$\tt 0^81........$}] (a081) at (0,-6) {};
    \node[b] (a0811) at (1,-6) {};
    \node[b] (a08111) at (2,-6) {};
    \node[b] (a081111) at (3,-6) {};
    \node[b] (a0811111) at (4,-6) {};
    \node[b] (a08111111) at (5,-6) {};
    \node[b,label=right:{$\tt 0^81^7..$}] (a0817) at (6,-6) {};

    \node[b,label=left:{$\tt 0^91.......$}] (a091) at (0,-7) {};
    \node[b] (a0911) at (1,-7) {};
    \node[b] (a09111) at (2,-7) {};
    \node[b] (a091111) at (3,-7) {};
    \node[b] (a0911111) at (4,-7) {};
    \node[b] (a09111111) at (5,-7) {};
    \node[b] (a0917) at (6,-7) {};
    \node[b,label=right:{$\tt 0^91^8$}] (a0918) at (7,-7) {};

    \node[b,label=left:{$\tt 0^{10}1......$}] (a0101) at (0,-8) {};
    \node[b] (a01011) at (1,-8) {};
    \node[b] (a010111) at (2,-8) {};
    \node[b] (a0101111) at (3,-8) {};
    \node[b] (a01011111) at (4,-8) {};
    \node[b] (a010111111) at (5,-8) {};
    \node[b,label=right:{$\tt 0^{10}1^7$}] (a01017) at (6,-8) {};

    \node[b,label=left:{$\tt 0^{11}1.....$}] (a011-1) at (0,-9) {};
    \node[b] (a011-11) at (1,-9) {};
    \node[b] (a011-111) at (2,-9) {};
    \node[b] (a011-1111) at (3,-9) {};
    \node[b] (a011-11111) at (4,-9) {};
    \node[b,label=right:{$\tt 0^{11}1^6$}] (a011-16) at (5,-9) {};

    \node[b,label=left:{$\tt 0^{12}1....$}] (a012-1) at (0,-10) {};
    \node[b] (a012-11) at (1,-10) {};
    \node[b] (a012-111) at (2,-10) {};
    \node[b] (a012-1111) at (3,-10) {};
    \node[b,label=right:{$\tt 0^{12}1^5$}] (a012-15) at (4,-10) {};

    \node[b,label=left:{$\tt 0^{13}1...$}] (a013-1) at (0,-11) {};
    \node[b] (a013-11) at (1,-11) {};
    \node[b] (a013-111) at (2,-11) {};
    \node[b,label=right:{$\tt 0^{13}1^4$}] (a013-14) at (3,-11) {};

    \node[b,label=left:{$\tt 0^{14}1..$}] (a014-1) at (0,-12) {};
    \node[b] (a014-11) at (1,-12) {};
    \node[b,label=right:{$\tt 0^{14}1^3$}] (a014-13) at (2,-12) {};

    \node[b,label=left:{$\tt 0^{15}1.$}] (a015-1) at (0,-13) {};
    \node[b,label=right:{$\tt 0^{15}1^2$}] (a015-12) at (1,-13) {};

    \node[b,label=left:{$\tt 0^{16}1$}] (a016-1) at (0,-14) {};
    
    \node[b,label=left:{$\tt 0^{17}$}] (a0) at (0,-15) {};
    \path[-]
    
    (a001) edge[looseness=0.7, bend right] (a0000001)
    (a00011) edge (a00001)
    (a00011) edge (a0001)

    (a000001111) edge (a00000111)
    (a00000111) edge (a00000011)
    (a00000011) edge (a071)

    (a071)      edge[sp, looseness=0.8, bend left] (a00001)
    (a071)      edge[looseness=0.8, bend left]     (a00001)

    ;

    \begin{scope}[on background layer]
      
      \path[fill=gray!30!white]
      (a0001.center)
      \foreach \i in {a00011, a00001}{ -- (\i.center) } -- cycle
      (a071.center)
      \foreach \i in {a00000111, a000001111, a081}{ -- (\i.center) } -- cycle
      (a011-1.center)
      \foreach \i in {a0711111, a07111111, a012-1}{ -- (\i.center) } -- cycle
      (a015-1.center)
      \foreach \i in {a0917, a0918, a016-1}{ -- (\i.center) } -- cycle;

      \path[-,draw]
      (a000001111) -- (a081)

      (a081) edge[bend right] (a011-1)
      (a011-1) -- (a0711111) --  (a0711111) -- (a07111111) -- (a07111111) -- (a012-1)
      (a012-1) edge[bend right] (a015-1)
      (a015-1) -- (a0917) -- (a0918) -- (a016-1)

      %

      (a016-1) edge[bend left] (a0)
      (a0) edge[sp, bend left, looseness=0.8]     (a013-1)
      (a0) edge[bend left, looseness=0.8]     (a013-1)
      (a013-1)  -- (a08111111) -- (a0817)

      (a0817) -- (a014-1)

      (a014-1) edge[sp, looseness=0.7, bend left] (a0101)
      (a014-1) edge[looseness=0.7, bend left] (a0101)

      (a0101) edge[double, double distance=1.5pt] (a071111)
      (a0911) -- (a091) -- (a0811) -- (a08111)
      
      (a071111) -- (a07111) -- (a0000001111) -- (a00000011111) -- (a071111)

      (a0000001) edge[double, double distance=1.5pt] (a0000011)
      (a0000011) edge (a0000111)
      (a0000111) edge (a000011)
      (a000011)  edge (a000001)
      (a000001)  edge (a0000011)

      (a0000001) edge[sp, looseness=0.7, bend right] (a0101)
      (a0000001) edge[looseness=0.7, bend right] (a0101)
      ;

    \end{scope}
  \end{tikzpicture}
  \caption{}
  \label{fig-big:b}
     \end{subfigure}

\caption{ (a) A decomposition of $\mathcal{Z}_{17}$ as a union of
  subsets $0^i1^j\cdot\mathcal{Z}_{17-i-j}$ (or equivalently a
  union of diagonals $\mathcal{D}^r_{17}$) and $\{0^{17}\}$. (b) An illustration of
  the 1-Gray code $\bb{Z}_{17}$. The pairs of consecutive
  diagonals dealt with Lemma 3 are shown in gray-filled area; the
  other pairs are dealt with Lemma 4. A point labelled $ \tt
  0^91.......$ (that is $ \tt 0^91$ followed by seven dots)
  corresponds to the set of words in $0^91\cdot\mathcal{Z}_{7}$.}
  
\label{fig-big}
\end{figure}

%% file: fig-big-rien.tex
{
  \centering
  \tikzset{
    b/.style={circle, fill, minimum size=1mm, inner sep=0pt, draw=white},
    sp/.style={line width=1mm, white},
    every label/.style={label distance=2mm}
  }
  \tt
  \scriptsize
  \begin{tikzpicture}[scale=0.5]
  
  \draw [->] (0,1) -- node[above] {$j$}  (4,1);
  \draw [->] (-1,0) -- node[left] {$i$}  (-1,-4);
  
    \node[b] (a001) at (0,0) {};

    \node[b] (a0001) at (0,-1) {};
    \node[b] (a00011) at (1,-1) {};

    \node[b] (a00001) at (0,-2) {};
    \node[b] (a000011) at (1,-2) {};
    \node[b] (a0000111) at (2,-2) {};

    \node[b] (a000001) at (0,-3) {};
    \node[b] (a0000011) at (1,-3) {};
    \node[b] (a00000111) at (2,-3) {};
    \node[b] (a000001111) at (3,-3) {};

    \node[b] (a0000001) at (0,-4) {};
    \node[b] (a00000011) at (1,-4) {};
    \node[b] (a000000111) at (2,-4) {};
    \node[b] (a0000001111) at (3,-4) {};
    \node[b] (a00000011111) at (4,-4) {};
    
    \node[b] (a071) at (0,-5) {};
    \node[b] (a0711) at (1,-5) {};
    \node[b] (a07111) at (2,-5) {};
    \node[b] (a071111) at (3,-5) {};
    \node[b] (a0711111) at (4,-5) {};
    \node[b] (a07111111) at (5,-5) {};
    
    \node[b] (a081) at (0,-6) {};
    \node[b] (a0811) at (1,-6) {};
    \node[b] (a08111) at (2,-6) {};
    \node[b] (a081111) at (3,-6) {};
    \node[b] (a0811111) at (4,-6) {};
    \node[b] (a08111111) at (5,-6) {};
    \node[b] (a0817) at (6,-6) {};

    \node[b] (a091) at (0,-7) {};
    \node[b] (a0911) at (1,-7) {};
    \node[b] (a09111) at (2,-7) {};
    \node[b] (a091111) at (3,-7) {};
    \node[b] (a0911111) at (4,-7) {};
    \node[b] (a09111111) at (5,-7) {};
    \node[b] (a0917) at (6,-7) {};
    \node[b] (a0918) at (7,-7) {};

    \node[b] (a0101) at (0,-8) {};
    \node[b] (a01011) at (1,-8) {};
    \node[b] (a010111) at (2,-8) {};
    \node[b] (a0101111) at (3,-8) {};
    \node[b] (a01011111) at (4,-8) {};
    \node[b] (a010111111) at (5,-8) {};
    \node[b] (a01017) at (6,-8) {};

    \node[b] (a011-1) at (0,-9) {};
    \node[b] (a011-11) at (1,-9) {};
    \node[b] (a011-111) at (2,-9) {};
    \node[b] (a011-1111) at (3,-9) {};
    \node[b] (a011-11111) at (4,-9) {};
    \node[b] (a011-16) at (5,-9) {};

    \node[b] (a012-1) at (0,-10) {};
    \node[b] (a012-11) at (1,-10) {};
    \node[b] (a012-111) at (2,-10) {};
    \node[b] (a012-1111) at (3,-10) {};
    \node[b] (a012-15) at (4,-10) {};

    \node[b] (a013-1) at (0,-11) {};
    \node[b] (a013-11) at (1,-11) {};
    \node[b] (a013-111) at (2,-11) {};
    \node[b] (a013-14) at (3,-11) {};

    \node[b] (a014-1) at (0,-12) {};
    \node[b] (a014-11) at (1,-12) {};
    \node[b] (a014-13) at (2,-12) {};

    \node[b] (a015-1) at (0,-13) {};
    \node[b] (a015-12) at (1,-13) {};

    \node[b] (a016-1) at (0,-14) {};
    
    \node[b,label=left:{$\tt 0^{17}$}] (a0) at (0,-15) {};
    
    \node  (truc) at (3.5, -1.) {$0^i1^j\cdot\mathcal{Z}_{17-i-j}$};
    \draw[<-] (a07111) -- (3, -1.3);

    \node (d1) at (-1, -11) {$\mathcal{D}^{14}_{17}$};
    \draw[-,dashed] (a08111111) -- (a013-1);
    
    \node (d1) at (-1, -12) {$\mathcal{D}^{15}_{17}$};
    \draw[-,dashed] (a0817) -- (a014-1);
  \end{tikzpicture}

}

%% file: fibograph.tex
\begin{figure}[!ht]
  \centering
  \tikzset{
    b/.style={circle, fill, minimum size=1.5mm, inner sep=0pt, draw=white},
    sp/.style={line width=1mm, white},
    every label/.style={label distance=0mm, fill=white}
  }
  \tt
  \scriptsize
  \tikzset{>=Stealth}
  \begin{tikzpicture}[scale=1]
    \node[b,label=below:000] (Z000) at (0,0) {};
    \node[b,label=below:100] (Z100) at (1,0) {};
    \path[->] (Z000) edge (Z100);
  \end{tikzpicture}
  \quad
  \begin{tikzpicture}[scale=1]
    \node[b,label=below:1000] (Z1000) at (0,0) {};
    \node[b,label=below:0000] (Z0000) at (1,0) {};
    \node[b,label=below:0100] (Z0100) at (2,0) {};
    \path[->] (Z1000) edge (Z0000);
    \path[->] (Z0000) edge (Z0100);
  \end{tikzpicture}
  \begin{tikzpicture}[scale=1]
    \node[b,label=below:00100] (Z00100) at (0,0) {};
    \node[b,label=below:00000] (Z00000) at (1,0) {};
    \node[b,label=below:01000] (Z01000) at (2,0) {};
    \node[b,label=above:10000] (Z10000) at (1,1) {};
    \node[b,label=above:11000] (Z11000) at (2,1) {};
    \path[->] (Z01000) edge (Z11000);
    \path[->] (Z11000) edge (Z10000);
    \path[->] (Z10000) edge (Z00000);
    \path[->] (Z00000) edge (Z00100);

    \path[-,densely dashed] (Z00000) edge (Z01000);
  \end{tikzpicture}\\
  \vspace{1em}
  \begin{tikzpicture}[scale=1]
    \node[b,label=below:100100] (Z100100) at (0,0) {};
    \node[b,label=below:100000] (Z100000) at (1,0) {};
    \node[b,label=below:110000] (Z110000) at (2,0) {};
    
    \node[b,label=above:000100] (Z000100) at (0,1) {};
    \node[b,label=above:000000] (Z000000) at (1,1) {};
    \node[b,label={[label distance=0mm]right:010000}] (Z010000) at (2,1) {};
    
    \node[b,label=above:001000] (Z001000) at (1,2) {};
    \node[b,label=above:011000] (Z011000) at (2,2) {};
    \begin{scope}[on background layer]
      
    \path[-, densely dashed] (Z000000) edge (Z001000) edge
    (Z010000);
    \path[-, densely dashed] (Z100100)  edge (Z100000);

    \path[->] (Z001000) edge (Z011000);
    \path[->] (Z011000) edge (Z010000);
    \path[->] (Z010000) edge (Z110000);
    \path[->] (Z110000) edge (Z100000);
    \path[->] (Z100000) edge (Z000000);
    \path[->] (Z000000) edge (Z000100);
    \path[->] (Z000100) edge (Z100100);
    
    \end{scope}
  \end{tikzpicture}
  \quad
  \begin{tikzpicture}[scale=1.5]
    \node[b,label=below:0011000] (Z0011000) at (0,0) {};
    \node[b,label=below:0010000] (Z0010000) at (1,0) {};
    \node[b,label=below:0110000] (Z0110000) at (2,0) {};
    \node[b,label=above:1110000] (Z1110000) at (2.5,0.5) {};
    \node[b,label=left:0001000] (Z0001000) at (0,1) {};    
    \node[b,label=above:0000000] (Z0000000) at (1,1) {};
    \node[b,label=above:0100000] (Z0100000) at (2,1) {};
    \node[b,label=above:1001000] (Z1001000) at (0.5, 1.5) {};
    \node[b,label=above:1000000] (Z1000000) at (1.5, 1.5) {};
    \node[b,label=above:1100000] (Z1100000) at (2.5, 1.5) {};
    \node[b,label=left:0000100] (Z0000100) at (1,2) {};
    \node[b,label=above:0100100] (Z0100100) at (2,2) {};
    \node[b,label=above:1000100] (Z1000100) at (1.5,2.5) {};

    \begin{scope}[on background layer]
      \path[-, densely dashed] (Z0000000)
      edge (Z0000100)
      edge (Z1000000)
      edge (Z0001000)
      ;
      \path[-, densely dashed] (Z0110000)
      edge (Z0010000)
      ;
      
      \path[-, densely dashed] (Z1001000)
      edge (Z1000000)
      ;

      \path[-, densely dashed] (Z1100000)
      edge (Z0100000)
      ;

      \path[-, densely dashed] (Z0100100)
      edge (Z0100000)
      ;

      \path[->] (Z1001000) edge (Z0001000);
      \path[->] (Z0001000) edge (Z0011000);
      \path[->] (Z0011000) edge (Z0010000);
      \path[->] (Z0010000) edge (Z0000000);
      \path[->] (Z0000000) edge (Z0100000);
      \path[->] (Z0100000) edge (Z0110000);
      \path[->] (Z0110000) edge (Z1110000);
      \path[->] (Z1110000) edge (Z1100000);
      \path[->] (Z1100000) edge (Z1000000);
      \path[->] (Z1000000) edge (Z1000100);
      \path[->] (Z1000100) edge (Z0000100);
      \path[->] (Z0000100) edge (Z0100100);

    \end{scope}
  \end{tikzpicture}
  \caption{Fibonacci-run graphs for small values of $n$.  Vertices
    correspond to the reverse of words from $\W^1_n$ beginning with
    $0$. The Hamiltonian path is provided by
    Corollary~\ref{conj-solved}. 
    If we read words from right to left, the path starts at
    ${0(001)^\star}$ and ends at ${(001)^\star}$.
}
  \label{fibograph}
\end{figure}
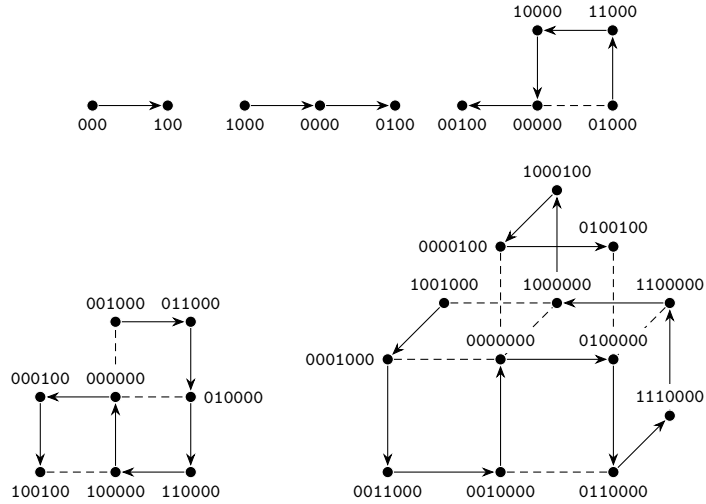

%% file: fig-sequence.tex
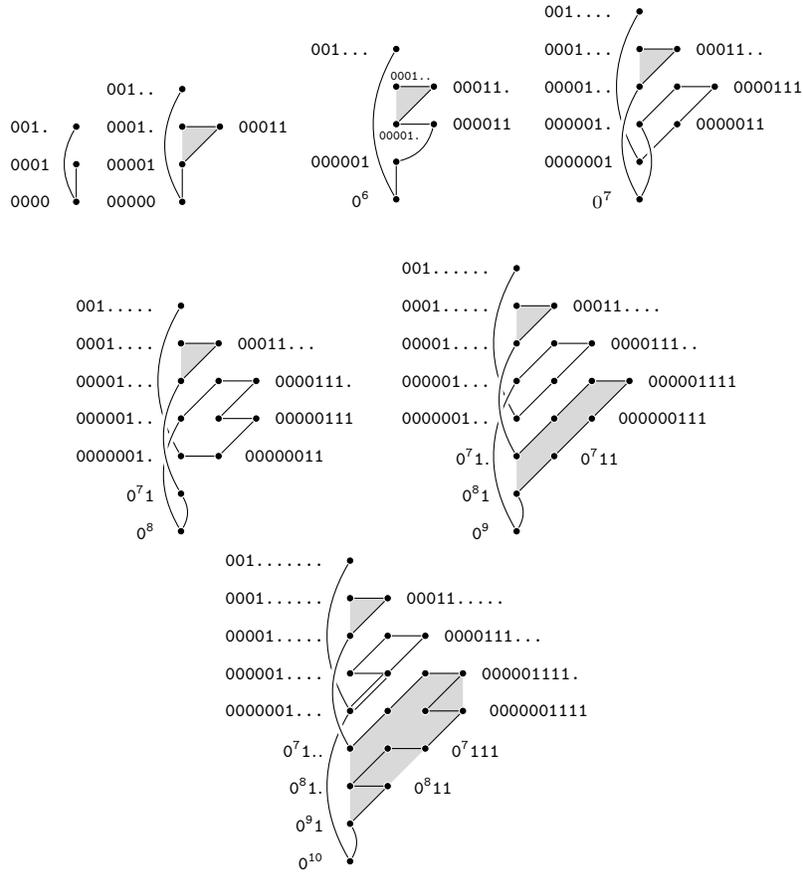
\begin{figure}[!ht]
  \centering
  \tikzset{
    b/.style={circle, fill, minimum size=1mm, inner sep=0pt, draw=white},
    sp/.style={line width=1mm, white},
    every label/.style={label distance=2mm}
  }
  \tt
  \scriptsize
  \begin{tikzpicture}[scale=0.5]
    \node[b,label=left:001.] (a001) at (0,0) {};
    \node[b,label=left:{0001}] (a0001) at (0,-1) {};
    \node[b,label=left:{0000}] (a0) at (0,-2) {};
    \path[-] (a001) edge[bend right] (a0)
    (a0001) edge (a0);
  \end{tikzpicture}\quad
  \begin{tikzpicture}[scale=0.5]
    \node[b,label={left:001..}] (a001) at (0,0) {};
    \node[b,label={left:0001.}] (a0001) at (0,-1) {};
    \node[b,label={[label distance=1mm]right:00011}] (a00011) at (1,-1) {};
    \node[b,label={left:00001}] (a00001) at (0,-2) {};
    \node[b,label={left:00000}] (a0) at (0,-3) {};
    \path[-] (a001) edge[bend right] (a0)
    (a0) edge (a00001)
    (a00001) edge (a00011)
    (a00011) edge (a0001);
    \begin{scope}[on background layer]
      \path[fill=gray!30!white]
      (a0001.center)
      \foreach \i in {a00011, a00001}{ -- (\i.center) } -- cycle;
    \end{scope}
  \end{tikzpicture}
  \begin{tikzpicture}[scale=0.5]
    \node[b,label=left:001...] (a001) at (0,0) {};
    \node[b,label={[label distance=1mm,xshift=2mm, yshift=-1mm, scale=0.7]above:0001..}] (a0001) at (0,-1) {};
    \node[b,label={[label distance=1mm]right:00011.}](a00011) at (1,-1) {};
    \node[b,label={[label distance=1mm,xshift=0.5mm, yshift=1mm, scale=0.7]below:00001.}] (a00001) at (0,-2) {};
    \node[b,label={[label distance=1mm]right:000011}] (a000011) at (1,-2) {};
    \node[b,label=left:{000001}] (a000001) at (0,-3) {};
    \node[b,label=left:{$\tt 0^6$}] (a0) at (0,-4) {};
    \path[-] (a001) edge[bend right] (a0)
    (a00011) edge (a00001)
    (a00011) edge (a0001)
    (a0) edge (a000001)
    (a000001) edge[bend right] (a000011)
    (a000011) edge (a00001)
    ;
    \begin{scope}[on background layer]
      \path[fill=gray!30!white]
      (a0001.center)
      \foreach \i in {a00011, a00001}{ -- (\i.center) } -- cycle;
    \end{scope}
  \end{tikzpicture}\quad
  \begin{tikzpicture}[scale=0.5]
    \node[b,label=left:001....] (a001) at (0,0) {};

    \node[b,label=left:0001...] (a0001) at (0,-1) {};
    \node[b,label={[label distance=1mm]right:00011..}] (a00011) at (1,-1) {};

    \node[b,label=left:00001..] (a00001) at (0,-2) {};
    \node[b] (a000011) at (1,-2) {};
    \node[b,label={[label distance=1mm]right:0000111}] (a0000111) at (2,-2) {};

    \node[b,label=left:{000001.}] (a000001) at (0,-3) {};
    \node[b,label={[label distance=1mm]right:0000011}] (a0000011) at (1,-3) {};

    \node[b,label=left:{0000001}] (a0000001) at (0,-4) {};

    \node[b,label=left:{$0^7$}] (a0) at (0,-5) {};

    \path[-]
    (a001) edge[bend right] (a0000001)
    (a00011) edge (a00001)
    (a00011) edge (a0001)

    (a0000001) edge (a0000011)
    (a0000011) edge (a0000111)
    (a0000111) edge (a000011)
    (a000011)  edge (a000001)

    (a000001) edge[sp, bend left] (a0)
    (a000001) edge[bend left] (a0)

    (a0)      edge[sp, bend left] (a00001)
    (a0)      edge[bend left] (a00001)
    
    ;
    \begin{scope}[on background layer]
      \path[fill=gray!30!white]
      (a0001.center)
      \foreach \i in {a00011, a00001}{ -- (\i.center) } -- cycle;
    \end{scope}
  \end{tikzpicture}
  
  \vspace{2em}
  \begin{tikzpicture}[scale=0.5]
    \node[b,label=left:\tt001.....] (a001) at (0,0) {};

    \node[b,label=left:\tt0001....] (a0001) at (0,-1) {};
    \node[b,label={[label distance=1mm]right:00011...}] (a00011) at (1,-1) {};

    \node[b,label=left:00001...] (a00001) at (0,-2) {};
    \node[b] (a000011) at (1,-2) {};
    \node[b,label={[label distance=1mm]right:0000111.}] (a0000111) at (2,-2) {};

    \node[b,label=left:{000001..}] (a000001) at (0,-3) {};
    \node[b] (a0000011) at (1,-3) {};
    \node[b,label={[label distance=1mm]right:00000111}] (a00000111) at (2,-3) {};

    \node[b,label=left:{0000001.}] (a0000001) at (0,-4) {};
    \node[b,label=right:{00000011}] (a00000011) at (1,-4) {};
    
    \node[b,label=left:{$\tt 0^71$}] (a071) at (0,-5) {};
    \node[b,label=left:{$\tt 0^8$}] (a0) at (0,-6) {};
    \path[-]
    (a001) edge[bend right] (a0000001)
    (a00011) edge (a00001)
    (a00011) edge (a0001)

    (a0000011)  edge (a0000111)
    (a0000011)  edge (a00000111)
    (a00000111) edge (a00000011)
    (a00000011) edge (a0000001)
    (a0000111)  edge (a000011)
    (a000011)   edge (a000001)

    (a071) edge[sp, bend left] (a0)
    (a071) edge[bend left]     (a0)

    (a000001) edge[sp, bend right] (a0)
    (a000001) edge[bend right] (a0)
    
    (a071)      edge[sp, bend left] (a00001)
    (a071)      edge[bend left]     (a00001)
    ;

    \begin{scope}[on background layer]
      \path[fill=gray!30!white]
      (a0001.center)
      \foreach \i in {a00011, a00001}{ -- (\i.center) } -- cycle;
    \end{scope}
  \end{tikzpicture}\quad\;
  \begin{tikzpicture}[scale=0.5]
    \node[b,label=left:\tt001......] (a001) at (0,0) {};

    \node[b,label=left:\tt0001.....] (a0001) at (0,-1) {};
    \node[b,label={[label distance=1mm]right:00011....}] (a00011) at (1,-1) {};

    \node[b,label=left:00001....] (a00001) at (0,-2) {};
    \node[b] (a000011) at (1,-2) {};
    \node[b,label={[label distance=1mm]right:0000111..}] (a0000111) at (2,-2) {};

    \node[b,label=left:{000001...}] (a000001) at (0,-3) {};
    \node[b] (a0000011) at (1,-3) {};
    \node[b] (a00000111) at (2,-3) {};
    \node[b,label={[label distance=1mm]right:000001111}] (a000001111) at (3,-3) {};

    \node[b,label=left:{0000001..}] (a0000001) at (0,-4) {};
    \node[b] (a00000011) at (1,-4) {};
    \node[b,label=right:{000000111}] (a000000111) at (2,-4) {};
    
    \node[b,label=left:{$\tt 0^71.$}] (a071) at (0,-5) {};
    \node[b,label=right:{$\tt 0^711$}] (a0711) at (1,-5) {};
    
    \node[b,label=left:{$\tt 0^81$}] (a081) at (0,-6) {};
    
    \node[b,label=left:{$\tt 0^9$}] (a0) at (0,-7) {};
    \path[-]
    (a001) edge[bend right] (a0000001)
    (a00011) edge (a00001)
    (a00011) edge (a0001)

    (a0000001) edge (a0000011)
    (a0000011) edge (a0000111)
    (a0000111) edge (a000011)
    (a000011)  edge (a000001)

    (a000001) edge[sp, bend right] (a0)
    (a000001) edge[bend right] (a0)

    (a081) edge (a0711)
    (a0711) edge (a000000111)
    (a000000111) edge (a000001111)
    (a000001111) edge (a00000111)
    (a00000111) edge (a00000011)
    (a00000011) edge (a071)

    (a071)      edge[sp, bend left] (a00001)
    (a071)      edge[bend left]     (a00001)

    (a0) edge[bend right] (a081)
    ;

    \begin{scope}[on background layer]
      \path[fill=gray!30!white]
      (a0001.center)
      \foreach \i in {a00011, a00001}{ -- (\i.center) } -- cycle
      (a071.center)
      \foreach \i in {a00000111, a000001111, a081}{ -- (\i.center) } -- cycle;
    \end{scope}
  \end{tikzpicture}\quad

  \begin{tikzpicture}[scale=0.5]
    \node[b,label=left:\tt001.......] (a001) at (0,0) {};

    \node[b,label=left:\tt0001......] (a0001) at (0,-1) {};
    \node[b,label={[label distance=1mm]right:00011.....}] (a00011) at (1,-1) {};

    \node[b,label=left:00001.....] (a00001) at (0,-2) {};
    \node[b] (a000011) at (1,-2) {};
    \node[b,label={[label distance=1mm]right:0000111...}] (a0000111) at (2,-2) {};

    \node[b,label=left:{000001....}] (a000001) at (0,-3) {};
    \node[b] (a0000011) at (1,-3) {};
    \node[b] (a00000111) at (2,-3) {};
    \node[b,label={[label distance=1mm]right:000001111.}] (a000001111) at (3,-3) {};

    \node[b,label=left:{0000001...}] (a0000001) at (0,-4) {};
    \node[b] (a00000011) at (1,-4) {};
    \node[b] (a000000111) at (2,-4) {};
    \node[b,label=right:{0000001111}] (a0000001111) at (3,-4) {};
    
    \node[b,label=left:{$\tt 0^71..$}] (a071) at (0,-5) {};
    \node[b] (a0711) at (1,-5) {};
    \node[b,label=right:{$\tt 0^7111$}] (a07111) at (2,-5) {};
    
    \node[b,label=left:{$\tt 0^81.$}] (a081) at (0,-6) {};
    \node[b,label=right:{$\tt 0^811$}] (a0811) at (1,-6) {};

    \node[b,label=left:{$\tt 0^91$}] (a091) at (0,-7) {};
    
    \node[b,label=left:{$\tt 0^{10}$}] (a0) at (0,-8) {};
    \path[-]
    (a001) edge[bend right] (a0000001)
    (a00011) edge (a00001)
    (a00011) edge (a0001)

    (a0000001) edge[sp, bend right] (a0)
    (a0000001) edge[bend right] (a0)

    (a000001111) edge (a00000111)
    (a00000111) edge (a00000011)
    (a00000011) edge (a071)

    (a000001111)  edge (a000000111)
    (a000000111)  edge (a0000001111)
    (a0000001111) edge (a07111)
    (a07111)      edge (a0711)
    (a0711)       edge (a081)
    (a081)        edge (a0811)
    (a0811)        edge (a091)

    (a071)      edge[sp, bend left] (a00001)
    (a071)      edge[bend left]     (a00001)

    (a091)  edge[bend left] (a0)
    ;

    \begin{scope}[on background layer]
      \path[]
      (a000011.center)
      \foreach \i in {a0000111, a0000001, a000001}{ -- (\i.center) } -- cycle;
    \end{scope}

    \path[-]
    (a0000001) edge[double, double distance=1.5pt] (a0000011)
    (a0000011) edge (a0000111)
    (a0000111) edge (a000011)
    (a000011)  edge (a000001)
    (a000001)  edge  (a0000011) 
    ;

    \begin{scope}[on background layer]
      \path[fill=gray!30!white]
      (a0001.center)
      \foreach \i in {a00011, a00001}{ -- (\i.center) } -- cycle
      
      (a071.center)
      \foreach \i in {a00000111, a000001111, 
        a0000001111,
        a091}{ -- (\i.center) } -- cycle;
    \end{scope}
  \end{tikzpicture}\quad
  \caption{An illustration of the recursive definition for the Gray
    codes $\bb{Z}_n$, $ 4 \le n \le 10 $.  A point labelled $ \tt
    001.......$ (that is $ \tt 0001$ followed by seven dots)
    corresponds to the set of words in $001\cdot\mathcal{Z}_{7}$.  }
  \label{fig-sequence}
\end{figure}